\documentclass{aastex}
\usepackage{emulateapj5}

\newcommand{\n}{\nodata}

\def\gtrsim{\mathrel{\hbox{\rlap{\hbox{\lower4pt\hbox{$\sim$}}}\hbox{$>$}}}}
\def\lesssim{\mathrel{\hbox{\rlap{\hbox{\lower4pt\hbox{$\sim$}}}\hbox{$<$}}}}

\slugcomment{Accepted for publication in the Astrophysical Journal}
\shortauthors{Lister et al.}
\shorttitle{Pearson-Readhead Survey From Space. I.}

\begin{document}
\title{The Pearson-Readhead Survey of Compact Extragalactic \\
 Radio Sources From Space. I. The Images}

\author{M. L. Lister\altaffilmark{1,2}, S. J. Tingay\altaffilmark{1,3},
D. W. Murphy\altaffilmark{1}, B. G. Piner\altaffilmark{1,4},
D. L. Jones\altaffilmark{1},\\ R. A. Preston\altaffilmark{1}}
\altaffiltext{1}{Jet Propulsion Laboratory, California Institute of Technology,
 MS 238-332, 4800 Oak Grove Drive, Pasadena, CA 91109-8099}
\altaffiltext{2}{Current address: NRAO, 520 Edgemont Road,
Charlottesville, VA 22903-2454}
\altaffiltext{3}{Australia Telescope National Facility, P.O. Box 76,
Epping, NSW 2121, Australia}
\altaffiltext{4}{Whittier College, Department of Physics, 13406
Philadelphia Street, Whittier, CA 90608}
\email{mlister@nrao.edu}

\begin{abstract}
We present images from a space-VLBI survey using the facilities of the
VLBI Space Observatory Programme (VSOP), drawing our sample from the
well-studied Pearson-Readhead survey of extragalactic radio sources.
Our survey has taken advantage of long space-VLBI baselines and large
arrays of ground antennas, such as the Very Long Baseline Array and
European VLBI Network, to obtain high resolution images of 27 active
galactic nuclei, and to measure the core brightness temperatures of
these sources more accurately than is possible from the ground. A
detailed analysis of the source properties is given in accompanying
papers. We have also performed an extensive series of simulations to
investigate the errors in VSOP images caused by the relatively
large holes in the $(u,v)$ plane when sources are observed near the orbit
normal direction. We find that while the nominal dynamic range
(defined as the ratio of map peak to off-source error) often exceeds
1000:1, the true dynamic range (map peak to on-source error) is only
about 30:1 for relatively complex core-jet sources. For sources
dominated by a strong point source, this value rises to approximately
100:1. We find the true dynamic range to be a relatively weak function
of the difference in position angle (PA) between the jet PA and
$(u,v)$ coverage major axis PA. For low signal-to-noise regions typically
located down the jet away from the core, large errors can occur,
causing spurious features in VSOP images that should be interpreted
with caution.
\end{abstract}


\section{INTRODUCTION}
Ground-based very long baseline interferometric (VLBI) surveys of the
parsec-scale regions of active galactic nuclei (AGNs) have led to
important insights into the nature of relativistic bulk outflows in
compact radio sources. Here we present observational data from the
first imaging survey of a complete sample of compact extragalactic
radio sources made with space-VLBI. Our objects are selected from the
well-studied Pearson-Readhead sample of AGNs, and are representative
of some of the brightest, most compact AGNs in the sky.

Space-VLBI represents a logical step in the evolution of radio
interferometry, and has the potential to provide the highest angular
resolution images of any observing technique. Since our understanding
of AGNs has historically been closely tied to resolution (e.g., the
discovery of jets with connected interferometers, and relativistic
bulk motions with VLBI), space-VLBI has an important role in AGN
research.

The first successful space-VLBI experiments were accomplished in 1986
and 1987 using a TDRSS communications satellite \citep{L86,LLU89,L90}, and
provided direct evidence for exceedingly compact emitting regions in
extragalactic radio sources. In 1997, the first satellite dedicated to
radio astronomical observations (HALCA) was launched by Japan's
Institute of Space and Astronautical Science, as part of the VLBI
Space Observatory Programme (VSOP). HALCA's highly elliptical orbit has
an apogee height of 21,400 km, making it possible to directly image
compact radio sources at 1.4 and 5 GHz with sub-milliarcsecond angular
resolution, and to measure brightness temperatures in excess of
$10^{12}$ K \citep{HHK98}.

The images presented here are of the highest angular resolution ever
achieved at 5 GHz, and reveal substantially more structural detail
than previous ground VLBI studies of the Pearson-Readhead sample
\citep[e.g.,][]{PR81, PR88, CWR93}. In \S2 we describe
our VSOP observations and data analysis methods, and present the
images and basic data on our sample. We describe our model fitting
technique in \S\ref{modelfitting}, and discuss the impact of
incomplete $(u,v)$ plane coverage on space-VLBI images in \S3. A full
analysis of our space-VLBI data is presented in two companion papers
(\citealt{TLP01}; \citealt{LPT01}).

\section{\label{analysis}OBSERVATIONS AND DATA ANALYSIS}
The Pearson-Readhead (PR) survey is a complete sample of 65 northern
($\delta > +35^o$) AGNs with 5 GHz flux densities $> 1.3$ Jy, and
galactic latitude $|b| > 10^o$ \citep{PR88}. In order to maximize the
likelihood of detecting fringes to HALCA (at the $7\sigma$ level), we
have selected as space-VLBI targets 31 PR objects that have flux
densities $> 0.4$ Jy on the longest Earth baselines ($\sim 10,000$
km).

Observations of 27 of these objects were conducted with the HALCA
orbiting antenna over the period 1997 August to 1999 April at a
frequency of 5 GHz, in conjunction with telescopes from either the
Very Long Baseline Array (VLBA), the European VLBI Network (EVN), or
in some cases, both arrays. Four sources (0454+844, 0804+499,
0945+408, and 0954+658) were not observed due to scheduling
constraints.  HALCA was tracked typically between four and five hours
per observation, synthesizing $(u,v)$ coverages with a maximum
baseline of nearly 500 M$\lambda$, or $~30,000$ km. A journal of our
observations is given in Table~\ref{obslog}.

Left circular polarization data from the ground antennas, or in the
case of the orbiter, the spacecraft tracking station, were recorded on
VLBA format tapes in two 16 MHz IFs and shipped to the VLBA correlator
in Socorro for processing.  The correlated data were
amplitude-calibrated, fringe-fitted, and frequency-averaged in
AIPS\footnote{AIPS and the VLBA are maintained by the National Radio
Astronomy Observatory, which is operated by Associated Universities,
Inc., under cooperative agreement with the National Science
Foundation.} before being exported to DIFMAP \citep{SPT94} for
self-calibration, imaging and model-fitting.

Once imported into DIFMAP, the visibilities were edited and then
coherently averaged over 30 second intervals. Due to its relatively
small diameter (8 m) and poor antenna efficiency, the system
equivalent flux density of HALCA is approximately 50 times worse than
that of a typical VLBA antenna at 5 GHz.  In order to give effectively
equal weight to the ground-only and ground-space visibilities, we
increased the weight of the HALCA antenna in self-calibration and
subsequent model-fitting by a factor based on the relative number of
ground-ground and space-ground visibilities and their respective
baseline sensitivities. These factors are listed in Table 1.

Following amplitude and phase self-calibration, we produced several
images for each source using different visibility weighting
schemes. The choice of weighting scheme has a large influence on
space-VLBI images, due to the unequal numbers of ground-ground and
ground-space visibilities, and the large differences in baseline noise
level described above. Uniform weighting reduces the influence of the
ground-ground visibilities on the image, which would normally dominate
since they outnumber the space-ground visibilities. In the right-hand
panels of Figures~1-9, we show uniformly-weighted images made
using all available baselines. These images have no additional
weighting by amplitude errors. In the left-hand panels, we show
naturally-weighted images made using ground-ground visibilities
only. We also include a representation of the restoring beam in the
lower left corner of each image, and a bar showing the spatial scale.
The latter is based on a standard Freidmann cosmology with zero
cosmological constant, deceleration parameter $q_o = 0.1$, and Hubble
constant $H_o = 100 \ h \ \rm km \ s^{-1} \ Mpc^{-1}$. We will assume
this model throughout this paper. The parameters of the restoring
beam, the peak brightness, and the contour levels of each image are
listed in Table~\ref{imageparameters}.

In Table~\ref{sourceprops} we list the basic data on each source,
including redshift, total amount of cleaned flux in the VLBI map, and
an estimate of the total (single-dish) 4.8 GHz flux density at the
epoch of the VSOP observation. The latter data were obtained by
interpolating data from the University of Michigan monitoring
program\footnote{http://www.astro.lsa.umich.edu/obs/radiotel/umrao.html},
and have an approximate error of $\pm 0.1$ Jy.

\subsection{Visibility Functions}
The visibility function of a source represents the observed power
measured on different spatial scales, with the long baselines
representing small-scale structure, and vice-versa. In
Figure~\ref{envelope} we plot the upper envelopes of the visibility
amplitude distribution versus projected baseline length for the
strongest and weakest sources, respectively. Instead of
dropping smoothly to zero, the curves show a pronounced flattening at
baselines greater than an earth diameter. This is indicative of
strong, compact components that remain unresolved at the longest
baselines. The brightness temperatures of these components and their
implications for relativistic beaming models are discussed more fully
by \cite{TLP01}.

\subsection{\label{modelfitting}Component Model-fitting}

We began our model-fitting analysis of each source with the core
component, which is commonly believed to be associated with a
flat-spectrum component located near the base of the jet. Wherever
possible, we relied on previously published identifications to
determine the position of the core. For sixteen sources the core
(labeled ``D'' in Figures~1-9) was located at the end of the
jet, and was also the brightest component in the 5 GHz image. In the
case of nine sources, the component at the end of the jet was
the brightest component in images at higher frequencies, but not at 5
GHz. The core component identifications for the two radio galaxies in
our sample are somewhat uncertain, as their jet structure may not be
one-sided.  For 3C~84 (0316+413), we arbitrarily assigned the core
position to the northernmost component of the jet, while for 2021+614,
we used the core identification of \cite{TSR00}.

We first removed the set of point source clean components used to fit
the core from the clean component model and replaced them with an
elliptical Gaussian described by 6 free parameters.  We then fit the
Gaussian in the $(u,v)$ plane using the MODELFIT task in DIFMAP,
which uses a linear least squares fit to the amplitude and phase to
find the best fit model.  For all our fits, we adjusted the visibility
weights used in the least squares calculation, such that the ground
antennas and the space antenna contributed equally to the reduced
chi-squared statistic.  We interspersed runs of MODELFIT with phase
self-calibration, to ensure that the reduced chi-squared was minimized
to find the best fit models. The parameters of our core-component fits
are given in Table~\ref{corecpts}.

After fitting the core, we removed the clean components associated
with various other isolated features and model-fit them with
elliptical Gaussians. In order to avoid introducing new nomenclature
for individual components, we have used existing names from the
literature wherever possible. These are given in Table~\ref{jetcpts},
along with the parameters of our fits.  We caution that these fits may
not be robust, especially in regions where there is continuous jet
emission. Indeed, some features (e.g., the southernmost component of
2021+614) were too complex to be represented by simple Gaussians, and
were not model-fitted.

We determined errors on the parameters of the core-component models
using the DIF\-WRAP package \citep{lov98}, a graphical interface to
DIFMAP which fixes given parameters of the models at different values
which are perturbations on the best-fit values, and allows all other
parameters to vary freely to re-minimize the reduced chi-squared.  In
this way it is possible to determine how far from the best-fit value a
given parameter can be forced before the model no longer fits the 
$(u,v)$ plane data, defining the error on that parameter.

For each source, we first determined whether the core component could
be fit with an unresolved point source (i.e., a delta-function
component). If a point source did not fit the visibilities within the
amplitude errors, we then tried one-dimensional (zero-axial ratio)
components of various lengths and position angles (PAs). If none of these
provided a good fit, we varied the parameters of the best-fit model as
described above to determine the range of possible brightness
temperature that still fit the visibility data.  In two cases
(0923+392 and 2021+614), the cores were too weak with respect to the
extended emission to determine a best fit range of brightness
temperature.

The components in our sample could have also been fit using the
surface brightness distributions of optically thin or optically thick
spheroids, instead of Gaussian profiles. This is due to the fact that
their Fourier transforms are virtually identical in the $(u,v)$ plane out
to the point at which the flux density is $50\%$ of the zero baseline
flux density \citep{TJP95}.  Since we do not generally observe past
this point for the core components in our sources, we have adopted
Gaussian surface brightness distributions.  To convert the observed
brightness temperature assuming a Gaussian component to an optically
thin or optically thick sphere, corrective factors of 0.67 and 0.56
respectively, should be applied.

We find that the sizes of our fitted components generally increase
exponentially with distance down the jet. In Figure~\ref{r__areacpt}
we present a log-log plot of fitted component area ($\Omega$) in
square parsecs against projected distance from the core component
($r$) in parsecs. A strong trend is present, with a linear correlation
coefficient of 0.42, and a significance level of $99.985\%$.  The
dashed line represents a linear regression through the points (in log
space) of the form $\log{\Omega} = (1.4 \pm 0.1)\log{r} + (0.07 \pm
0.1)$. The fit considering quasars alone is $\log{\Omega} = (1.1 \pm
0.2)\log{r} + (0.5
\pm 0.2)$, and for BL Lacertae objects alone it is $\log{\Omega} = (1.5
\pm 0.2)\log{r} - (0.2 \pm 0.2)$. These fitted slopes are all
significantly less than two, which is the value for a conical jet of
constant opening angle. Considering the fact that the projected
distances in Fig.~\ref{r__areacpt} are likely to be severely
foreshortened due to small viewing angles (see Paper II), the trend
indicates that the jets are still in a state of expansion and are
undergoing collimation at distances of several hundred parsecs from
the core.

\subsection{Influence of $(u,v)$ Coverage on Model-fits}
The $(u,v)$ coverages of most of our space-VLBI observations are
highly elongated, due to the elliptical orbit of the HALCA satellite
(Figure~\ref{uvplot}). As a result, the
angular extent of the jets and components in our sample are much
better constrained along one direction. We have examined the possible
effects of this unequal $(u,v)$ plane sampling on our core component
model-fits by inter-comparing the position angles of the fitted
Gaussian component major axis, the position angle of the restoring
beam, and the position angle of the parsec-scale jet.

We find that the major axis position angle of the fitted core
component is strongly correlated with that of the beam and
$(u,v)$ coverage, and therefore does not likely reflect an intrinsic
property of the source. For those sources whose cores could be fit
with a zero-axial ratio (ZAR) component, the $(u,v)$ coverage is
generally perpendicular to the jet direction, and the fitted major
axis of the ZAR component is aligned with the jet. This suggests that
there are two or more closely spaced components in the core region
that are separated by much less than a beam width. Since they cannot
be resolved, the visibility data can be fit with a long, narrow (ZAR)
component. In this case, the upper limit on the brightness temperature
is unconstrained.

\section{\label{uvcoverage} EFFECTS OF INCOMPLETE $(U,V)$ PLANE
COVERAGE ON SPACE-VLBI IMAGES} 

Most of the space-VLBI observations of our sample were scheduled when
the source lay close to the orbit normal.  Under these conditions the
maximum possible spatial resolution can be achieved, at the cost of
creating relatively large holes in the $(u,v)$ plane aperture
coverage. This is due to the the fact that the HALCA spacecraft has a
highly elliptical orbit with an apogee height of 21,400 km, which
greatly exceeds an Earth diameter.  The effects of $(u,v)$ plane
under-sampling are well-documented for ground-based inteferometric
arrays \citep[e.g.,][]{P89}, but relatively little work has been
published for the case of space-VLBI. Here we discuss the effects of
$(u,v)$-holes on the quality of our space-VLBI images, and how the
resulting imaging errors depend on the source's jet position angle
relative the position angle of the major axis of the $(u,v)$ coverage.

\subsection{Example Simulations}

We have addressed these issues by performing extensive image
simulations using several different source models. These simulations
have thermal amplitude and phases errors added to them at the same
level as is present in the real data.  Thus, the RMS noise-level on
simulated HALCA baselines is about 7 times higher than on the
simulated ground-only baselines.  The simulated data were imaged using
the same DIFMAP package that was used to reduce our actual space-VLBI
data.  For the purposes of this discussion we will consider the image
artifacts which are present in 5 GHz images for the following source
models, all of which are composed of Gaussian components:

1) A complex core-jet source which has a total flux
density of 2.7 Jy. We will refer to this model as {\it cj-complex};

2) A simple core-jet source, which has a 1 Jy, 10$^{12}$ K core with
axial ratio unity and a 1 Jy, 10$^{11}$ K `jet' with axial ratio 1/3
whose centroid is located at a distance of 1.21 milliarcseconds away
from the core. The major axis of this component measures 1.21
milliarcseconds, and is aligned with the core-jet direction. We will
refer to this model as {\it cj-simple};

3) A 1 Jy,  10$^{12}$ K circular Gaussian;

4) A 1 Jy, 10$^{14}$ K circular Gaussian. This source is completely
unresolved for VSOP 5 GHz observations and can be used to determined
the image artifacts in unresolved point sources. It is therefore a useful
point of reference when examining the artifacts in more complex
sources.


In Figures~\ref{fig_diffex1} and \ref{fig_diffex2} we show two
simulations we have performed using the {\it cj-complex} and
{\it cj-simple} models respectively. We use the same $(u,v)$ coverage
in both cases (panel b), which has a major axis PA of
0$^{\circ}$. Both jets have a PA of 60$^{\circ}$.

Panels (a), (c), and (d) show the model, simulated image, and
difference image, respectively, all convolved with an identical beam
that corresponds to uniform weighting of the visibility data. The
difference image is simply defined as the difference between the the
simulated image and the model. The difference images reveal a common
error feature seen in the imaging of core-jet sources, in the form of
a sinusoidal ripple that runs along the length of the jet. This
feature is the result of a well-known instability of the CLEAN
algorithm, and has modulations that tend to correspond with the
spacing of  holes in the $(u,v)$ plane \citep{C83}.  In our
simulations, this error feature is so prominent that it can cause
visible artifacts to occur in the jet that may easily be
over-interpreted as discrete components. It is important, therefore,
to exercise caution in the interpretation of relatively weak features
in VSOP images.

A good example of this can be found in the model in Figure
\ref{fig_diffex1}, in which panel (a) shows a rather smooth jet. The
simulated VSOP image, on the other hand, shows a series of discrete
components which run down the jet (panel c). In panels (d) and (f) the
difference map signal-to-noise ratio (SNR) at each pixel and the
percentage error in the simulated are shown as a function of the
simulated image SNR. The SNRs quoted here use a reference noise level
determined far away from the source structure.  From these two panels
we can see how the imaging errors decrease as the map SNR
increases. For high SNR points ($>$100) the image error is typically
less than $10\%$. However, for intermediate SNR points ($10 < {\rm
SNR} < 100$) the percentage error can almost be 100\%, so caution must
be exercised in over-interpreting data with this SNR.

\subsection{Dynamic Range}

A common method of quantifying image errors is to take the
maximum in the image divided by the off-source RMS noise level
($DR_{\rm RMS}$), which is sometimes referred to as the dynamic range
of an image. For example, the simulated images shown in Figures
\ref{fig_diffex1} and \ref{fig_diffex2} have  $DR_{\rm RMS}$ values of
1600:1 and 2800:1 respectively. However, as our simulations show, this
is not the true dynamic range in the image as the on-source errors are
much larger than the off-source errors. We define here a new dynamic
range, called $DR_{\rm diff}$ which is the maximum in the image
divided by the maximum value (whether positive or negative) in the
difference image. Using this definition, $DR_{\rm diff}$ has values of
34:1 and 28:1 for the images in Figures \ref{fig_diffex1} and
\ref{fig_diffex2}. Consequently the true dynamic ranges ($DR_{\rm diff}$)
of these images are a {\it factor of $\sim 50$ worse} compared to the dynamic
range using the off-source noise level ($DR_{\rm RMS}$).  For
comparison,  the images made using the single $10^{12}$ K and $10^{14}$
K Gaussian component models, have values of $DR_{\rm RMS}$ of 3000:1
and 4000:1, and $DR_{\rm diff}$ of 48:1 and 130:0, respectively. Thus,
we see a general trend of increasing image fidelity as the source
structure becomes simpler (i.e., $DR_{\rm diff}$ decreases). However,
even in the limit of a unresolved point source, the value of
the true dynamic range is limited to approximately $100:1$.

\subsection{Effects of Jet Position Angle on Image Dynamic Range}

One might expect that the dynamic range that we can obtain on a given
observation of a source should depend on the difference between a
source's jet PA and the PA of the $(u,v)$ coverage major
axis ($\Delta PA$). For $\Delta PA$ =0$^{\circ}$ we get the highest
resolution along the jet and for $\Delta PA$ = 90$^{\circ}$ we get the
highest resolution perpendicular to the jet.  Thus, the ability of the
sinusoidal image error to develop could depend on the value of $\Delta
PA$. Consequently, certain alignments of jet position angle with
respect to the $(u,v)$ coverage might be more reliable than those of a
different $\Delta PA$. We have investigated this phenomenon by
conducting a series of simulations using the same $(u,v)$ coverage as
shown in Figures \ref{fig_diffex1} and
\ref{fig_diffex2} but for different values of the jet PA in the model sources
{\it cj-complex} and {\it cj-simple}. In Figures \ref{fig_pa1} and
\ref{fig_pa2} we plot how both $DR_{\rm RMS}$ and $DR_{\rm diff}$ vary
as a function of the jet PA while keeping the PA of the major axis of
the $(u,v)$ coverage fixed. As can be seen, $DR_{\rm RMS}$ is almost
independent of $\Delta PA$ and $DR_{\rm diff}$ is only a relatively
weak function of this position angle difference.

\section{SUMMARY}

We have presented data from the first imaging survey of a complete
sample of compact extragalactic radio sources made with
space-VLBI. These data represent the highest angular resolution images
ever obtained for these objects at 5 GHz, and provide important
information regarding the structure of AGN jets on parsec scales. In
particular, we have found an exponential trend of increasing component
size with distance down the jet, which indicates that these jets are
still undergoing expansion on scales of several hundred parsecs. We
have also detected a pronounced flattening in the visibility functions
of our sources, which is indicative of strong, compact components that
remain unresolved at the longest space-Earth baselines. We discuss the
implications of this trend and the other general properties of our
sample in two companion papers (\citealt{TLP01}; \citealt{LPT01}).

We have performed an extensive series of simulations to investigate
the image errors in VSOP images caused by the relatively large holes
in the $(u,v)$ plane when sources are observed near the orbit normal
direction. We find that while the nominal dynamic range ($DR_{\rm
RMS}$) often exceeds 1000:1, the true dynamic range ($DR_{\rm diff}$)
is about 30:1 for relatively complex core-jet sources. For sources
dominated by a strong point source, this value rises to approximately
100:1. The true dynamic range is also found to be a relatively weak
function of the difference in position angle between the jet PA and
$(u,v)$ coverage major axis PA. For high SNR regions in the image
(SNR$>$100) the error on individual pixels is typically less than
$10\%$. However, for low SNR regions typically located down the jet
away from the core, large errors can occur and spurious features can be
seen. Caution should therefore be exercised when interpreting regions of
low SNR in VSOP images. 

\acknowledgments
We gratefully acknowledge the VSOP Project, which is led by the
Japanese Institute of Space and Astronautical Science in cooperation
with many organizations and radio telescopes around the world. This
research has made use of data from the University of Michigan Radio
Astronomy Observatory, which is supported by the National Science
Foundation and by funds from the University of Michigan. This work was
undertaken in part at the Jet Propulsion Laboratory, California Institute of
Technology, under contract with the National Aeronautics and Space
Administration.  SJT acknowledges support through an NRC/NASA-JPL
Research Associateship.

\begin{deluxetable}{llllllr}

\tablecolumns{7}
\tablewidth{0pt}
\tablecaption{\label{obslog}Journal of Observations}
\tablehead{ \colhead{IAU} &  \colhead{Other} & \colhead{Observing} & 
\colhead{Ground}&  \colhead{Observing} & \colhead{Integration}    &
\colhead{HALCA}\\ 
\colhead{Name} &     \colhead{Name} &   \colhead{Date} &
\colhead{Antenna Array}  & \colhead{Freq.[GHz]} &\colhead{Time
[h]\tablenotemark{a}} & \colhead{Weight\tablenotemark{b}} }
\startdata
0016+731&  \n      & 1998 Mar 2  & VLBA+U & 4.81 &10 (3) & 1156 \\
0133+476& OC 457   & 1999 Aug 16 & VLBA  & 4.81  &5  (4.5)  &   274 \\
0153+744&  \n      & 1998 Aug 18 & VLBA$-$Br,Sc,La    &4.81& 6 (5.5)  & 187 \\
0212+735&  \n      & 1997 Sep 5&  VLBA$-$Sc & 4.97 & 8 (4.5) &     415  \\
0316+413& 3C 84    & 1998 Aug 25& VLBA+Y & 4.97 & 6 (5)    &  538 \\
0711+356& OI 318   & 1999 Apr 9& VLBA   & 4.81 & 6 (4) &  324  \\
0814+425& OJ 425   & 1999 Apr 24&VLBA  & 4.81 &  6  (5)   & 289  \\
0836+710& 4C 71.07 & 1997 Oct 7&VLBA    & 4.97  &  5.5  (3.5)   &  390  \\
0859+470& 4C 47.29 & 1999 Feb 14& JEGNMTWO & 4.97 &6.5  (1)  &  1460\\
0906+430& 3C 216   & 1999 Feb 14& JEGNMTW & 4.97 &  5.5 (3)   & 399 \\
0923+392& 4C 39.25 & 1997 Oct 23& VLBA+Y & 4.97 & 2.5 (2)    & 847 \\
1624+416& 4C 41.32 & 1998 Feb 7&  VLBA$-$Mk & 4.81 & 4 (1.5)   & 347  \\
1633+382& 4C 38.41 & 1998 Aug 4& VLBA+Y & 4.81 & 5.5 (4)     & 531  \\
1637+574& OS 562   & 1998 Apr 21& VLBA$-$Sc,Ov,Kp & 4.81 &6.5 (4.5) &280  \\
1641+399& 3C 345   & 1998 Jul 28& VLBA+EY & 4.81 & 4.5 (3.5)  & 674  \\
1642+690& 4C 69.21 & 1998 May 31& JEGNMWS & 4.97 & 6.5 (4.5)  &   225  \\
1652+398& MK 501   & 1998 Apr 7& VLBA+E & 4.81 & 5 (3.5)  & 370  \\
1739+522& 4C 51.37 & 1998 Jun 14& VLBA & 4.81  & 6.5 (4) &  391  \\
1749+701&  \n      & 1999 Jun 1& VLBA & 4.81 &  8  (5)  &  406  \\
1803+784&  \n      & 1997 Oct 17& VLBA & 4.97 &  6 (4)  &  370 \\
1807+698& 3C 371   & 1998 Mar 11& VLBA & 4.81 &  6 (5)  &  349  \\
1823+568& 4C 56.27 & 1998 May 31& JEGNMO & 4.97 & 4.5 (3.5)   &   542\\
1828+487& 3C 380   & 1998 Jul 4& VLBA+E & 4.81  & 6 (4)   &  584  \\
1928+738& 4C 73.18 & 1997 Aug 22& VLBA+E & 4.97 & 5.5 (4.5)  &  482 \\
1954+513& OV 591   & 1997 Nov 10& EGNMWO & 4.97 &  12 (4)  &  777 \\ 
2021+614& OW 637   & 1997 Nov 7& VLBA+ENMO & 4.97 & 5.5 (4.5)  &  575  \\
2200+420& BL Lac   & 1997 Dec 8& VLBA &    4.97 &  6.5 (5) & 315    \\
\enddata

\tablecomments{Antenna abbreviations: Br = Brewster, E = Effelsberg, G = Green
Bank, J=Jodrell Bank, Kp = Kitt Peak, La = Los Alamos, M = Medicina, N
= Noto, O = Onsala, Ov = Owens Valley, S = Sheshan, Sc = Saint Croix,
T = Torun, U = Usuda, W = Westerbork, Y = Phased Very Large Array. \\ }
\tablenotetext{a}{Approximate ground integration time on source. Values in
parentheses indicate integration time that included the HALCA
spacecraft.}
\tablenotetext{b}{Weighting assigned to the HALCA antenna during
self-calibration and model-fitting (ground visibilities had a nominal
weighting factor of unity).}
\end{deluxetable}

\begin{deluxetable}{llrrrl}
\tabletypesize{\footnotesize}
\tablecolumns{6}
\tablecaption{\label{imageparameters}Summary of Image parameters}
\tablewidth{0pt}
\tablehead{ 
\colhead{ Source} & \colhead{Baselines} &
\colhead{ Beam}&  \colhead{ PA} &
  \colhead{ Peak} &
\colhead{ Contour levels}\\
\colhead{ (1)} & \colhead{ (2)} & \colhead{ (3)}  &
\colhead{ (4)} & \colhead{ (5)} &
\colhead{ (6)}}

\startdata
0016+731 & Ground & 1.37 x 2.02 & $-18$ & 0.450 & $-0.06$, \ldots, 61.44 \\
  & All & 0.38 x 0.77 & $-76$ & 0.216 & $-1.5$, \ldots, 96 \\
0133+476 & Ground & 1.72 x 4.09 & 3 & 1.795& $-0.1$, \ldots, 51.2\\
 & All &0.23 x 0.61 & $-29$ & 1.340 &  $-0.5$, \ldots, 64\\
0153+744 & Ground & 1.03 x 2.10 & $-30$ & 0.295& $-1$, \ldots, 64\\
  & All &  0.36 x 0.52 & $-36$  & 0.070 &   $-4$, \ldots, 64\\
0212+735 & Ground &1.65 x 2.94 & 76 & 2.357   & $-0.1$, \ldots, 51.2 \\
 & All & 0.24 x 0.67 & 58 & 0.786 & $-0.4$, \ldots, 51.2 \\
0316+413 & Ground & 1.27 x 2.80 & $-26$ & 1.944 &$-1.5$, \ldots, 96 \\
& All & 0.31 x 0.77 & 0 & 0.375 & $-2.5$, \ldots, 80 \\
0711+356 & Ground & 1.56 x 3.26 & $-4$ & 0.707 & $-0.2$, \ldots, 51.2\\
& All & 0.31 x 0.55 & $-63$ & 0.189 & $-4$, \ldots, 64 \\
0814+425 & Ground & 1.55 x 3.11 & $-21$ & 0.589   & $-0.2$, \ldots, 51.2 \\
& All & 0.26 x 0.82 & $-42$ & 0.437 & $-0.75$, \ldots, 96 \\
0836+710 & Ground & 1.57 x 2.88 & 51 & 0.844& $-0.25$, \ldots, 64
 \\
& All & 0.22 x 0.62 & $-36$ & 0.372 & $-1.25$, \ldots, 80 \\
0859+470 & Ground & 0.93 x 2.20 & $-19$ & 0.533 & $-0.5$, \ldots, 64 \\
& All & 0.20 x 1.88 & $-32$ & 0.264 & $-5$, \ldots, 80 \\
0906+430 & Ground &2.23 x 5.70 & $-30$ & 0.610 & $-0.65$, \ldots, 83.2 \\
& All &0.17 x 0.82 & $-44$ & 0.381& $-1.5$, \ldots, 96\\
0923+392 & Ground &1.67 x 3.63 & $-16$ & 9.490 & $-0.1$, \ldots, 51.2\\
& All &0.28 x 2.28 & $-18$ & 4.092 & $-0.5$, \ldots, 64 \\
1624+416 & Ground &1.90 x 4.46 & $-83$ & 0.419& $-0.6$, \ldots, 76.8 \\
& All & 0.22 x 1.08 & 20 & 0.126 & $-4.5$, \ldots, 72 \\
1633+382 & Ground & 1.37 x 3.44 & 5 & 0.893 & $-0.1$, \ldots, 51.2, 95 \\
& All & 0.25 x 0.66 & $-27$ &0.397 & $-0.75$, \ldots, 96 \\
1637+574 & Ground & 1.55 x 2.97 & $-28$ & 0.506 & $-0.25$, \ldots, 64 \\
& All & 0.26 x 0.52 & 4 & 0.226 & $-1.5$, \ldots, 96 \\
1641+399 & Ground & 0.62 x 3.15 & $-8$ & 1.877 & $-0.2$, \ldots, 51.2 \\
& All & 0.24 x 0.64 & $-14$ & 0.940 & $-1$, \ldots, 64 \\
1642+690 & Ground & 2.17 x 2.83 & 8 & 0.625 & $-0.25$, \ldots, 64 \\
& All & 0.22 x 0.47 & $-10$ & 0.372 & $-1.25$, \ldots, 80 \\
1652+398 & Ground & 1.40 x 2.93 & $-42$ & 0.509 & $-0.4$, \ldots, 51.2 \\
& All & 0.25 x 0.62 & $31$ &0.325 & $-1$, \ldots, 64 \\
1739+522 & Ground & 1.62 x 3.11 & $-2$ & 1.814 & $-0.04$, \ldots, 81.92 \\
& All & 0.22 x 0.58 & $-4$ & 0.918 & $-0.5$, \ldots, 64 \\
1749+701 & Ground & 1.83 x 2.98 & 72 & 0.312 & $-0.05$, \ldots, 51.2 \\
& All & 0.25 x 0.77 & 75 & 0.169 & $-1.75$, \ldots, 56 \\
1803+784 & Ground & 1.61 x 2.95 & $-29$ & 1.539 & $-0.125$, \ldots, 64 \\
& All & 0.22 x 0.58 & $-82$ & 0.928 & $-0.8$, \ldots, 51.2 \\
1807+698 & Ground & 1.61 x 2.87 & $-50$ & 0.567 & $-0.07$, \ldots, 71.68 \\
& All & 0.28 x 0.46 & 66 & 0.320 & $-0.1$, \ldots, 51.2 \\
1823+568 & Ground & 1.66 x 2.51 & $-6$ & 0.817 & $-0.3$, \ldots, 76.8 \\
& All & 0.25 x 0.68 & 16 & 0.536 & $-0.75$, \ldots, 96 \\
1828+487 & Ground & 0.95 x 1.67 & $-40$ & 0.631 & $-0.28$, \ldots, 71.68 \\
& All & 0.27 x 0.62 & $-9$ & 0.379  &$-0.9$, \ldots, 57.6 \\
1928+738 & Ground & 0.76 x 1.59 & 23 & 1.311 & $-0.3$, \ldots, 76.8 \\
& All & 0.40 x 0.54 & $-42$ & 0.918 & $-0.4$, \ldots, 51.2 \\
1954+513 & Ground & 1.30 x 2.54 & $-3$ & 0.600 & $-0.2$, \ldots, 51.2 \\
& All & 0.25 x 0.48 & $-34$ & 0.228 & $-4$, \ldots, 64 \\
2021+614 & Ground & 0.99 x 1.47 & 66 & 1.112 & $-0.4$, \ldots, 51.2 \\
& All & 0.23 x 0.44 & $-51$ & 0.401 & $-1.2$, \ldots, 76.8 \\
2200+420 & Ground & 1.57 x 3.05 & $-24$ & 2.265 & $-0.55$, \ldots, 70.4 \\
& All & 0.22 x 0.46 & $-27$ & 1.239 & $-1$, \ldots, 64 \\

\enddata
\tablecomments{Columns are as follows: (1) Source
name; (2) VLBI baselines used in image; (3) FWHM dimensions of
Gaussian restoring beam, in milliarcseconds; (4) Position angle of
restoring beam, in degrees; (5) Peak intensity of image [$\rm Jy \
beam^{-1}$]; (6) Minimum and maximum contour levels, expressed as a
percentage of peak intensity (intermediate contours are separated by
factors of two).}

\end{deluxetable}

\begin{deluxetable}{lllrrr}

\tablecolumns{7}
\tablewidth{0pt}
\tablecaption{\label{sourceprops}Source Properties}
\tablehead{ \colhead{IAU} &  \colhead{Other} & \colhead{Opt.} & &
\colhead{$S_{VLBI}$} &\colhead{$S_{tot}$} \\
\colhead{Name} &     \colhead{Name} &   \colhead{Id.} & \colhead{z} 
& \colhead{[Jy]} &\colhead{[Jy]} \\
\colhead{(1)} & \colhead{(2)} & \colhead{(3)} & \colhead{(4)} &
\colhead{(5)} & \colhead{(6)} }

\startdata
0016+731& \n        &Q&1.781 &0.581& 0.8 \\  
0133+476&OC 457     &Q&0.859 &1.985& 2.0     \\  
0153+744&\n         &Q&2.338 &1.090& 1.1 \\  
0212+735&\n         &Q&2.367 &3.045& 3.0     \\  
0316+413&3C 84      &RG&0.017 &17.494&22.2 \\  
0711+356&OI 318     &Q&1.620 &1.030& 1.0 \\  
0814+425&OJ 425     &BL&0.245 &0.836& 1.0 \\  
0836+710&4C 71.07   &Q&2.180 &1.762& 2.2 \\  
0859+470&4C 47.29   &Q&1.462 &0.974& 1.3 \\  
0906+430&3C 216     &Q&0.670 &0.684& 1.6 \\  
0923+392&4C 39.25   &Q&0.699 &10.968&11.0 \\  
1624+416&4C 41.32   &Q&2.550 &0.668& 0.9 \\  
1633+382&4C 38.41   &Q&1.807 &1.889& 2.4 \\  
1637+574&OS 562     &Q&0.749 &0.703& 0.9 \\  
1641+399&3C 345     &Q&0.595 &5.971& 8.2 \\  
1642+690&4C 69.21   &Q&0.751 &0.878& 1.2 \\  
1652+398&MK 501     &BL&0.033 &0.977& 1.7 \\  
1739+522&4C 51.37   &Q&1.381 &1.933& 2.2 \\  
1749+701&\n         &BL&0.770 &0.504& 0.6  \\  
1803+784&\n         &BL&0.680 &2.040& 2.2 \\  
1807+698&3C 371     &BL&0.050 &0.880& 1.6 \\  
1823+568&4C 56.27   &BL&0.663 &1.047& 1.6 \\  
1828+487&3C 380     &Q&0.692 &1.957& 5.3 \\  
1928+738&4C 73.18   &Q&0.302 &3.324& 3.7 \\  
1954+513&OV 591     &Q&1.223 &0.848& 1.4 \\  
2021+614&OW 637     &RG&0.228 &2.664& 2.8 \\  
2200+420&BL Lac     &BL&0.069 &4.069& 4.4 \\  
\enddata

\tablecomments{Columns are as follows: (1) IAU Name; (2) Alternate
name; (3) Optical identification, where Q = quasar, BL = BL Lacertae
object, RG = radio galaxy; (4) Redshift; (5) Total 5 GHz cleaned flux
density in VLBI image in Janskys; (6) Single dish 5 GHz flux density
at VLBI observation epoch in Janskys, estimated from the University of
Michigan light curve. }

\end{deluxetable}

\begin{deluxetable}{lrrrrll}

\tablecolumns{7}
\tablewidth{0pt}
\tablecaption{\label{corecpts}Core Component Properties}
\tablehead{ \colhead{Source} 
& \colhead{S}   & \colhead{Maj.} & \colhead{Axial} &
\colhead{PA}  &\colhead{Fit} &\colhead{$T_b$}  \\
    \colhead{Name} &  
  \colhead{[Jy]} & \colhead{axis} & \colhead{ratio} &
\colhead{[deg.]} & \colhead{Type}& \colhead{$[/10^{12}\;\rm K]$} \\
\colhead{(1)} & \colhead{(2)} & \colhead{(3)} & \colhead{(4)} & \colhead{(5)} &
\colhead{(6)} & \colhead{(7)} }
\startdata
0016+731  &  0.118  & 0.53&0.45  &$-$64 & ZAR & $0.05^{+\inf}_{-0.03}$   \\[4pt]
0133+476  &  1.554 &0.17 & 0.70 & $-$6 & ZAR & $3.9^{+\inf}_{-1.0}$   \\[4pt]
0153+744 &   0.352 & 1.17 & 0.40 & $-$86 &Res & $0.034^{+0.002}_{-0.01}$ \\[4pt]
0212+735 &   1.206 & 0.54 & 0.72 & 71 & Res & $0.28^{+0.06}_{-0.03}$ \\[4pt]
0316+413 &   1.556 & 1.17 & 0.87 & 62 & Res &$0.066^{+0.01}_{-0.02}$ \\[4pt]
0711+356 &  0.078 & 0.46 & 0.19 & $-$36 & ZAR & $0.10^{+\inf}_{-0.07}$ \\[4pt]
0814+425 &  0.448 & 0.09 & 0.53 & $-$66 & U & $5.4^{+\inf}_{-4.5}$ \\[4pt]
0836+710 &  0.603 & 0.28 & 0.46 & 49 & ZAR & $0.83^{+\inf}_{-0.4}$ \\[4pt]
0859+470 &  0.447 & 0.85 & 0.16 & $-$31 & Res & $0.19^{+0.05}_{-0.05}$ \\[4pt]
0906+430 &  0.483 & 0.43 & 0.15 & $-$34 & Res & $0.86^{+0.8}_{-0.2}$ \\[4pt]
0923+392 &  0.267 & 1.01 & 0.34 & $-$21 & n/a\tablenotemark{a} & $0.038$ \\[4pt]
1624+416 &  0.152 & 0.37 & 0.29 & 31 & U & $0.20^{+\inf}_{-0.1}$ \\[4pt]
1633+382 &  0.443 & 0.15 & 0.71 & $-$43 & Res & $1.46^{+0.4}_{-0.2}$ \\[4pt]
1637+574 &  0.241 & 0.13 & 0.24 & 27 & U & $3.13^{+\inf}_{-0.3}$ \\[4pt]
1641+399 &  0.811 & 0.40 & 0.63 & $-$26 & Res & $0.42^{+0.05}_{-0.01}$ \\[4pt]
1642+690 &  0.396 & 0.19 & 0.14 & $-$22 & ZAR & $3.87^{+\inf}_{-1.6}$ \\[4pt]
1652+398 &  0.450 & 0.24 & 0.87 & 32 & Res & $0.47^{+0.4}_{-0.1}$ \\[4pt]
1739+522 &  1.745 & 0.33 & 0.88 & $-$29 & Res & $0.96^{+0.03}_{-0.03}$ \\[4pt]
1749+701 &  0.264  & 0.32 & 0.52 & $-$50 & ZAR & $0.26^{+\inf}_{-0.06}$ \\[4pt]
1803+784 &  1.374 & 0.29 & 0.63 & 88 & Res & $1.28^{+0.6}_{-0.2}$ \\[4pt]
1807+698 &  0.387 & 0.28 & 0.29 & 71 & ZAR & $0.90^{+1.1}_{-0.3}$ \\[4pt]
1823+568 &  0.791 & 0.60 & 0.20 & 16 & Res & $0.54^{+0.1}_{-0.1}$ \\[4pt]
1828+487 &  0.510 & 0.42 & 0.12 & $-32$ & ZAR & $1.3^{+\inf}_{-0.5}$ \\[4pt]
1928+738 &  0.416 & 0.35 & 0.34 & $-$21 & ZAR & $0.49^{+\inf}_{-0.4}$ \\[4pt]
1954+513 &  0.313 & 0.40 & 0.23 & $-$38 & ZAR & $0.42^{+\inf}_{-0.1}$ \\[4pt]
2021+614 &  0.078 & 0.70 & 0.40 & 60 & n/a\tablenotemark{a} & $0.02$ \\[4pt]
2200+420 &  1.552 & 0.42 & 0.44 & 10 & Res & $0.99^{+0.3}_{-0.2}$ \\[4pt]
\enddata
\tablecomments{Columns are as follows: (1) Source name;  (2) Fitted 5
GHz flux density in Janskys; (3) Major axis of fitted Gaussian (FWHM)
in milliarcseconds; (4) Axial ratio of component; (5) Position angle
of component's major axis; (6) Visibility data consistent within
errors with: U = unresolved component; ZAR = component having
zero-axial ratio; Res = component that is resolved on longest space
baselines; (7) Brightness temperature of best component fit in units
of $10^{12} K$, with range of possible values that fit the visibility
data.}

\tablenotetext{a}{Component too weak to determine errors in model fit parameters.}
\end{deluxetable}

\begin{deluxetable}{lrrrcrrrc} 
\tablecolumns{9} 
\tabletypesize{\footnotesize} 
\tablewidth{0pt}  
\tablecaption{\label{jetcpts}Jet Component Properties}  
\tablehead{\colhead{Cpt.} & \colhead {$r$} &  
\colhead{$\theta$} & \colhead{S} & \colhead{Maj.} &  
\colhead{Axial} & \colhead{PA} & \colhead{log $T_b$} & \\  
\colhead{Name} & \colhead{[mas]} & \colhead{[deg.]} & \colhead{[Jy]} & 
\colhead{Axis} & \colhead{Ratio} & \colhead{[deg.]} & \colhead{[K]} & \colhead{Ref.} \\  
\colhead{(1)} & \colhead{(2)} & \colhead{(3)} & \colhead{(4)} &  
\colhead{(5)} & \colhead{(6)} & \colhead{(7)} & \colhead{(8)} & \colhead{[9]}}   
\startdata 
\sidehead{0016+731} 
C2 &  0.65 &   130 &0.254 & 0.72 & 0.37 &   83&    10.9  & \n \\* 
C1 &  0.89 &   137 &0.183 & 0.95 & 0.35 &  -64&    10.5  & \n \\* 
\sidehead{0133+476} 
C3 &  0.63 &   -31 &0.221 & 0.89 & 0.16 &    0&    11.0  & \n \\* 
C2 &  1.44 &   -32 &0.102 & 0.88 & 0.44 &  -64&    10.2  & \n \\* 
C1 &  3.75 &   -45 &0.078 & 3.95 & 0.73 &  -44&     8.6  & \n \\* 
\sidehead{0153+744} 
B  & 10.16 &   157 &0.592 & 1.91 & 0.58 &  -62&    10.2  & 1 \\* 
\sidehead{0212+735} 
C6 &  0.34 &   124 &0.880 & 0.32 & 0.37 &   -7&    12.1  & \n \\* 
C5 &  0.60 &   117 &0.347 & 0.42 & 0.00 &  -22&    \n  & \n \\* 
C4 &  2.21 &   106 &0.347 & 1.68 & 0.45 &   88&    10.1  & \n \\* 
C3 &  4.55 &   107 &0.047 & 1.70 & 0.29 &   89&     9.4  & \n \\* 
C2 &  7.00 &   103 &0.050 & 1.55 & 0.41 &  -52&     9.4  & \n \\* 
C1 & 14.06 &    92 &0.179 & 2.40 & 0.68 &  -53&     9.4  & \n \\* 
\sidehead{0316+413} 
C2 &  1.07 &  -134 &0.501 & 0.96 & 0.36 &   21&    10.9  & \n \\* 
C1 &  1.30 &  -164 &1.109 & 0.84 & 0.67 &   43&    11.1  & \n \\* 
\sidehead{0711+356} 
C2 &  0.90 &   -20 &0.061 & 0.61 & 0.79 &  -45&    10.0  & \n \\* 
C1 &  5.80 &   -23 &0.764 & 0.83 & 0.76 &  -53&    10.9  & \n \\* 
\sidehead{0814+425} 
C4 &  0.36 &   100 &0.064 & 0.27 & 0.22 &  -31&    11.3  & \n \\* 
C3 &  1.05 &    87 &0.140 & 0.51 & 0.79 &  -18&    10.6  & \n \\* 
C2 &  1.40 &    84 &0.104 & 0.44 & 0.58 &  -56&    10.7  & \n \\* 
C1 &  2.27 &   140 &0.047 & 2.10 & 0.44 &    8&     9.1  & \n \\* 
\sidehead{0836+710} 
C4 &  0.40 &  -132 &0.097 & 0.17 & 0.18 &   79&    11.9  & \n \\* 
C3 &  1.29 &  -138 &0.143 & 0.61 & 0.47 &  -30&    10.6  & \n \\* 
C2 &  2.68 &  -143 &0.204 & 0.52 & 0.46 &  -20&    10.9  & \n \\* 
C1 & 11.57 &  -148 &0.178 & 1.65 & 0.76 &   -0&     9.6  & \n \\* 
\sidehead{0859+470} 
C2 &  1.73 &    -9 &0.366 & 1.82 & 0.74 &  -43&     9.9  & \n \\* 
C1 &  4.26 &     6 &0.161 & 7.07 & 0.16 &  -10&     9.0  & \n \\* 
\sidehead{0906+430} 
C3 &  0.40 &   167 &0.075 & 0.22 & 0.32 &  -45&    11.4  & \n \\* 
C2 &  2.16 &   147 &0.050 & 1.78 & 0.13 &  -51&     9.8  & 2  \\* 
B  &  4.63 &   153 &0.035 & 0.85 & 0.24 &  -81&    10.0  & 2 \\* 
\sidehead{0923+392} 
C4 &  0.51 &   110 &0.263 & 1.69 & 0.21 &  -16&    10.4  & \n \\* 
C3 &  1.92 &    97 &2.670 & 1.29 & 0.75 &   48&    11.1  & \n \\* 
C2 &  2.04 &   100 &6.536 & 0.43 & 0.68 &   74&    12.4  & \n \\* 
C1 &  2.48 &    96 &1.259 & 0.51 & 0.48 &  -31&    11.7  & \n \\* 
\sidehead{1624+416} 
C3 &  0.49 &  -119 &0.084 & 0.86 & 0.32 &    4&    10.3  & \n \\* 
C2 &  1.26 &  -124 &0.147 & 0.93 & 0.26 &   44&    10.5  & \n \\* 
C1 &  2.69 &  -143 &0.114 & 2.57 & 0.54 &   32&     9.2  & \n \\* 
\sidehead{1633+382} 
C5 &  0.50 &   -78 &0.270 & 0.47 & 0.15 &  -59&    11.6  & \n \\* 
C4 &  1.02 &   -75 &0.442 & 1.20 & 0.26 &  -34&    10.8  & \n \\* 
A &  1.91 &   -86 &0.599 & 0.84 & 0.47 &  -29&    11.0  & 3 \\* 
C2 &  3.00 &   -88 &0.046 & 0.54 & 0.56 &   -6&    10.2  & \n \\* 
C1 &  4.06 &   -73 &0.035 & 1.60 & 0.11 &  -25&     9.8  & \n \\* 
\sidehead{1637+574} 
C2 &  0.97 &  -157 &0.303 & 0.49 & 0.55 &    4&    11.1  & \n \\* 
C1 &  2.90 &  -160 &0.098 & 1.14 & 0.72 &   49&     9.7  & \n \\* 
\sidehead{1641+399} 
C9 &  0.64 &  -122 &1.891 & 0.39 & 0.74 &   49&    11.9  & 4 \\* 
C8 &  1.22 &  -102 &1.569 & 0.40 & 0.92 &   49&    11.8  & 4 \\* 
C7 &  2.33 &  -101 &0.607 & 0.63 & 0.86 &  -86&    11.0  & 4 \\* 
\\
\\
\\
\\
\\
\\
\sidehead{1642+690} 
C4 &  0.48 &  -178 &0.086 & 0.28 & 0.54 &   16&    11.0  & \n \\* 
C3 &  1.52 &  -172 &0.143 & 0.51 & 0.53 &   11&    10.7  & \n \\* 
C2 &  2.64 &  -159 &0.107 & 0.40 & 0.58 &   -1&    10.7  & \n \\* 
C1 &  8.39 &  -164 &0.051 & 1.97 & 0.36 &  -88&     9.3  & \n \\* 
\sidehead{1652+398} 
C1 &  0.89 &   168 &0.067 & 0.49 & 0.26 &   63&    10.8  & \n \\* 
\sidehead{1739+522} 
C1 &  0.41 &    38 &0.078 & 0.48 & 0.26 &  -12&    10.8  & \n \\* 
\sidehead{1749+701} 
C3 &  1.23 &   -72 &0.082 & 0.95 & 0.35 &   82&    10.1  & \n \\* 
C2 &  2.12 &   -74 &0.034 & 0.84 & 0.52 &   31&     9.7  & \n \\* 
C1 &  2.99 &   -62 &0.033 & 1.02 & 0.63 &    3&     9.4  & \n \\* 
\sidehead{1803+784} 
C1 &  1.38 &   -98 &0.413 & 0.65 & 0.82 &   50&    10.8  & \n \\* 
\sidehead{1807+698} 
C1 &  0.66 &  -102 &0.231 & 0.74 & 0.28 &  -85&    10.9  & \n \\* 
\sidehead{1823+568} 
C3 &  1.24 &  -157 &0.123 & 0.60 & 0.23 &   42&    10.9  & \n \\* 
C2 &  3.25 &  -165 &0.060 & 1.88 & 0.14 &    1&     9.8  & \n \\* 
C1 &  7.39 &  -163 &0.024 & 0.58 & 0.69 &   41&     9.7  & \n \\* 
\sidehead{1828+487} 
C2 &  3.28 &   -28 &0.170 & 0.60 & 0.73 &   18&    10.5  & \n \\* 
A  &  9.26 &   -31 &0.165 & 0.69 & 0.74 &  -87&    10.4  & 5 \\* 
\sidehead{1928+738} 
C5 &  0.67 &   149 &1.292 & 0.43 & 0.42 &  -14&    11.9  & 6 \\* 
C4 &  1.95 &   159 &0.429 & 0.60 & 0.61 &    7&    11.0  & 6 \\* 
C3 &  2.31 &   173 &0.622 & 0.80 & 0.46 &  -22&    11.0  & \n \\* 
C2 &  3.68 &   160 &0.190 & 1.34 & 0.42 &   -4&    10.1  & 6 \\* 
C1 &  5.22 &   164 &0.051 & 1.06 & 0.38 &  -79&     9.8  & 6 \\* 
\sidehead{1954+513} 
C1 &  0.64 &   -56 &0.348 & 0.43 & 0.37 &  -27&    11.4  & \n \\* 
\sidehead{2021+614} 
B  &  3.89 &    34 &0.639 & 0.71 & 0.58 &  -47&    11.0  & 7\\* 
C3 &  4.28 &    33 &0.143 & 0.36 & 0.78 &   73&    10.8  & \n \\* 
C2 &  4.74 &    34 &0.136 & 0.94 & 0.24 &  -41&    10.5  & \n \\* 
A  &  6.77 &    46 &0.101 & 1.32 & 0.28 &  -61&    10.0  & 7 \\* 
\sidehead{2200+420} 
C3 &  0.88 &  -171 &0.323 & 0.55 & 0.47 &   14&    11.1  & \n \\* 
C2 &  1.35 &  -160 &0.805 & 0.43 & 0.96 &   83&    11.3  & \n \\* 
C1 &  2.43 &  -169 &0.929 & 1.59 & 0.50 &   11&    10.6  & \n \\* 
\enddata 
\tablecomments{Columns are as follows: (1) Component name; (2) Distance 
from core in milliarcseconds; (3) Position angle with respect to 
core; (4) Flux density in Jy; (5) Major axis of fitted component in 
milliarcseconds; (6) Axial ratio of fitted component; (7) Position 
angle of component's major axis; (8) log of Gaussian brightness 
temperature in K; (9) Reference for component identifications:
(1) \citealt{HKW97}; (2) \citealt{PFF00}; (3) \citealt{LS00}; (4) \citealt{KZR00};
(5) \citealt{KSI00}; (6) \citealt{MPP00}; (7) \citealt{TSR00}. }
\end{deluxetable}

\begin{figure}
\plotone{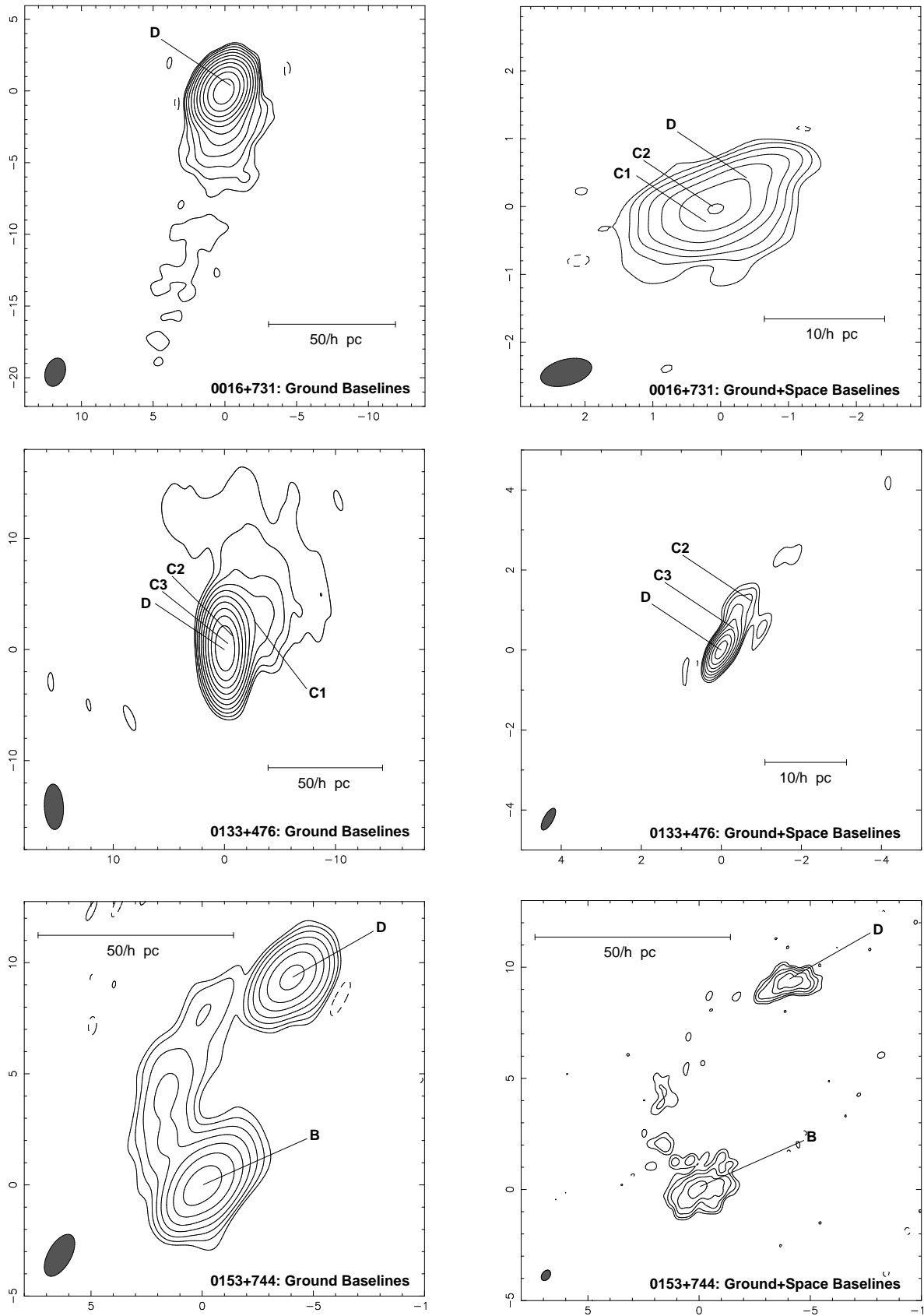}
\caption{\label{images}Left column: naturally-weighted images using
ground-to-ground baselines only. Right column: uniformly-weighted images using
all baselines. Angular scales are in milliarcseconds. }
\end{figure}

\newpage
\begin{figure}
\plotone{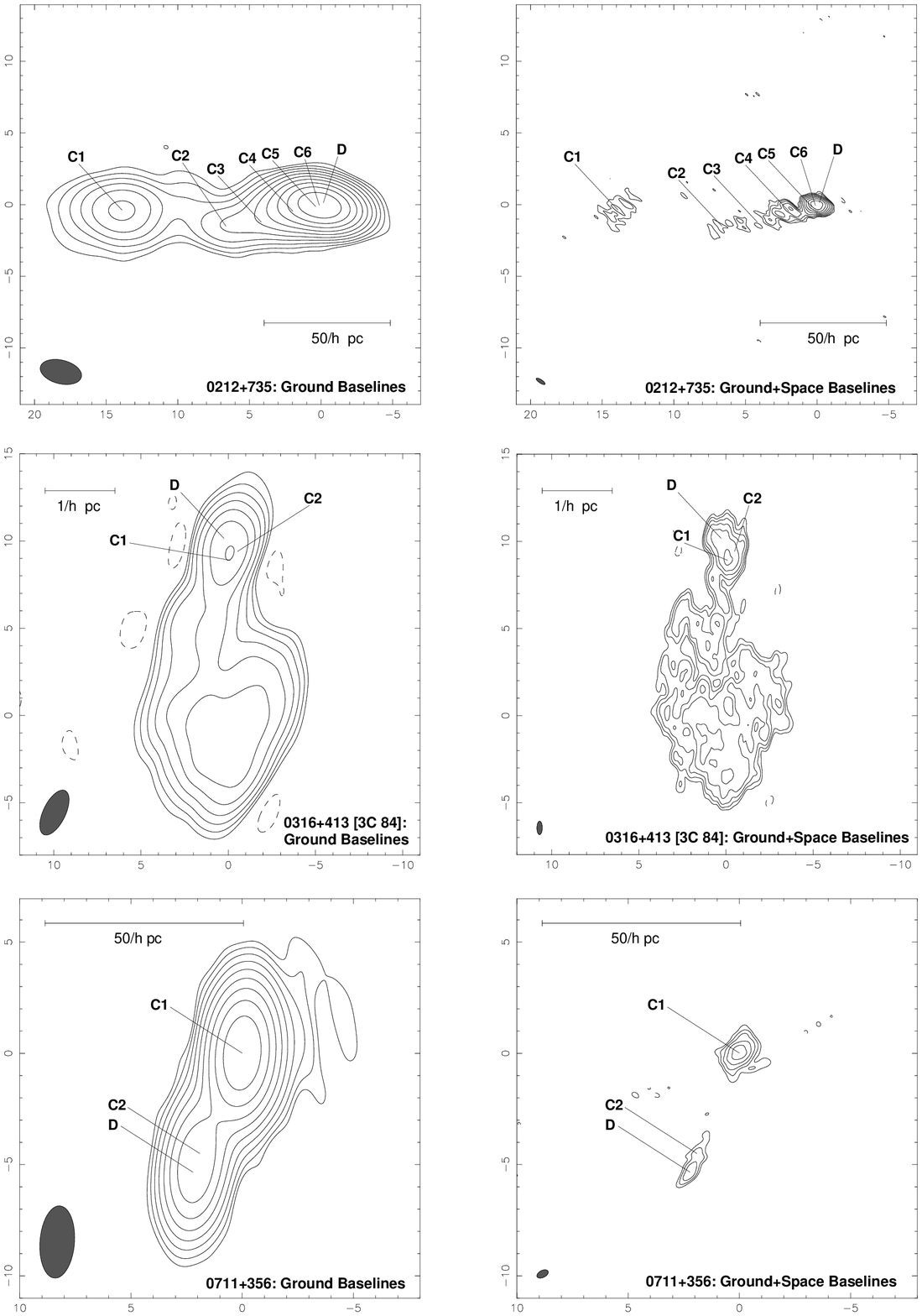}
\caption{\small Left column: naturally-weighted images using
ground-to-ground baselines only. Right column: uniformly-weighted images using
all baselines. Angular scales are in milliarcseconds. }
\end{figure}

\newpage

\begin{figure}
\plotone{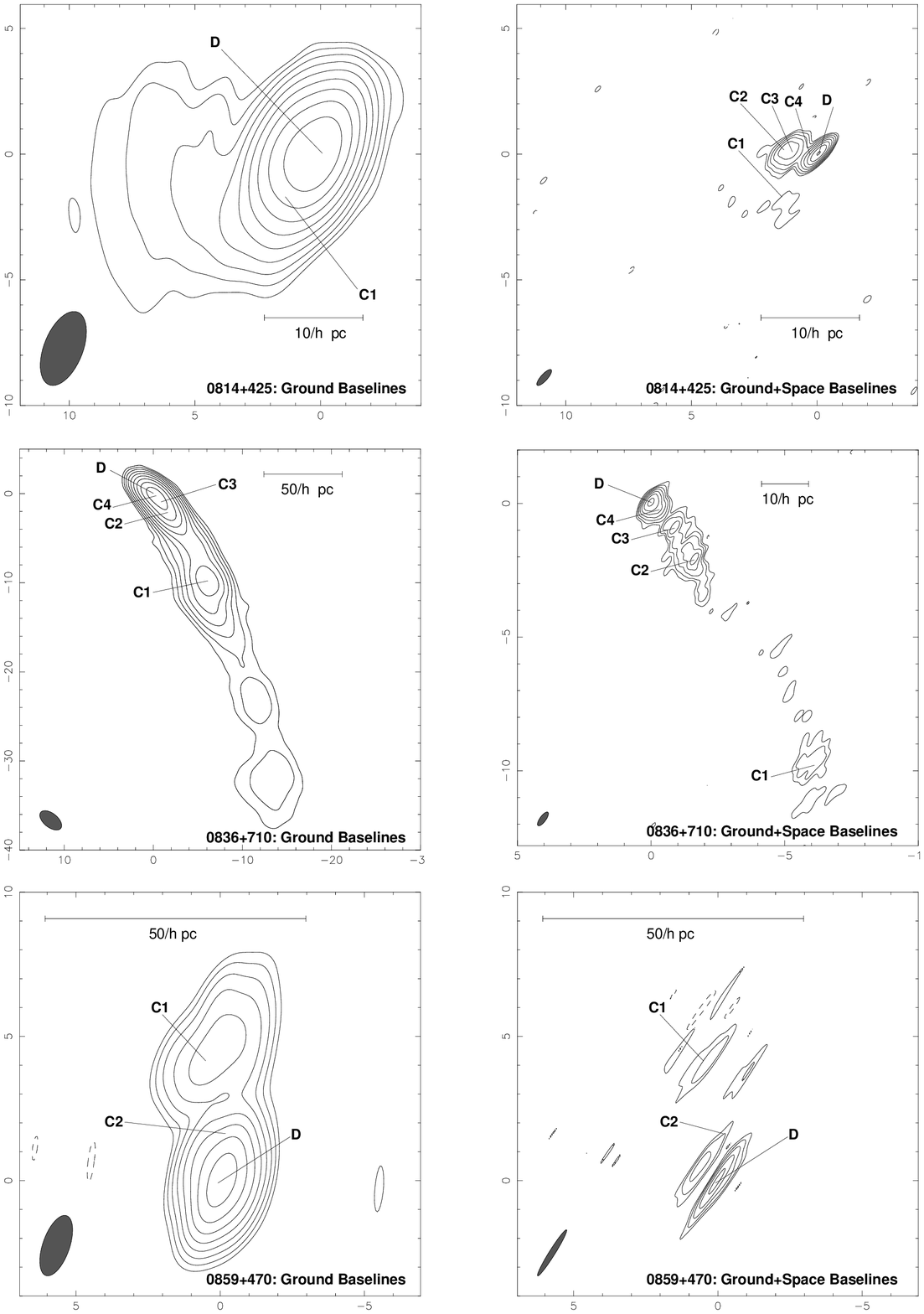}
\caption{\small Left column: naturally-weighted images using
ground-to-ground baselines only. Right column: uniformly-weighted images using
all baselines. Angular scales are in milliarcseconds. }
\end{figure}

\begin{figure}
\plotone{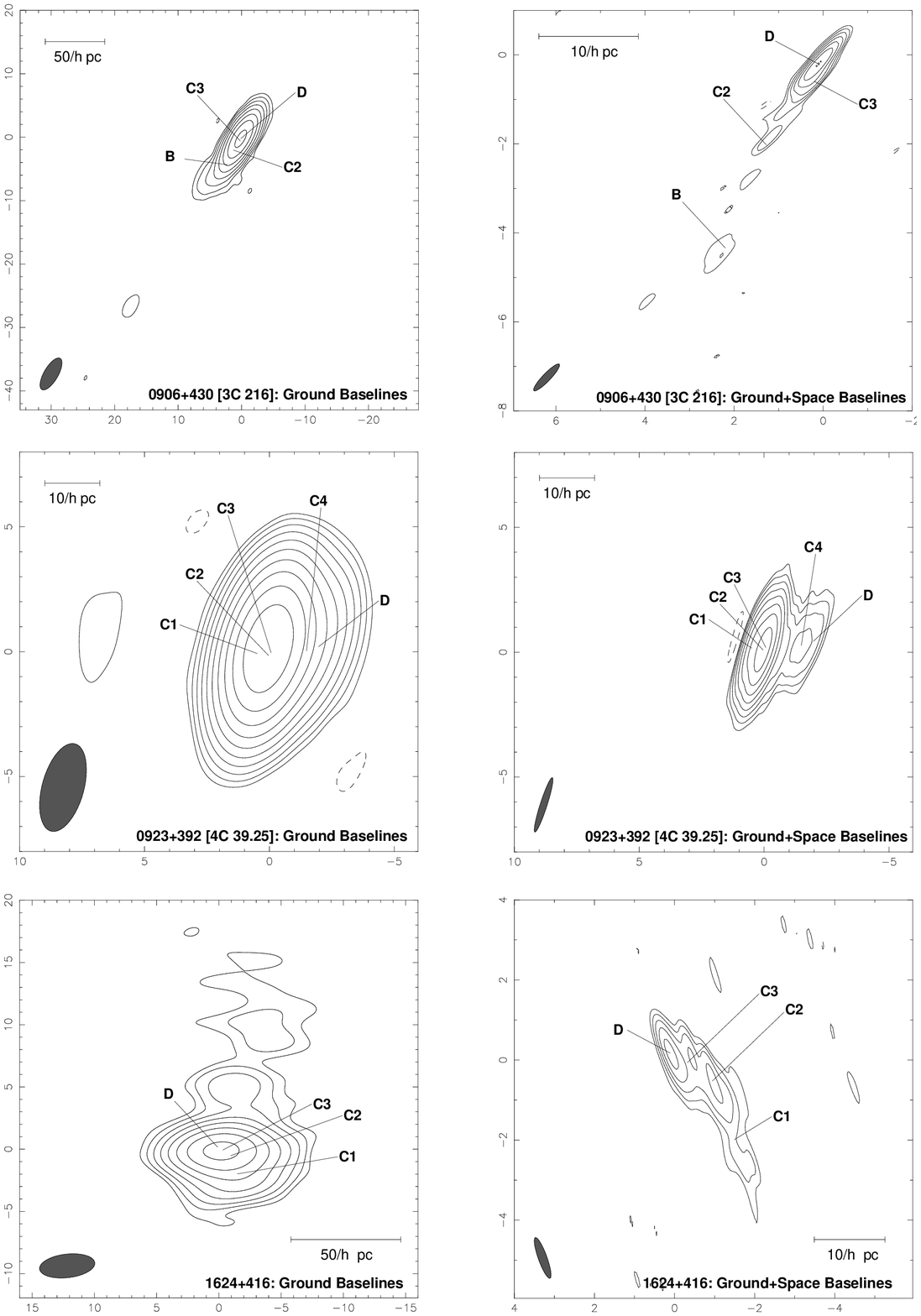}
\caption{\small Left column: naturally-weighted images using
ground-to-ground baselines only. Right column: uniformly-weighted images using
all baselines. Angular scales are in milliarcseconds. }
\end{figure}

\begin{figure}
\plotone{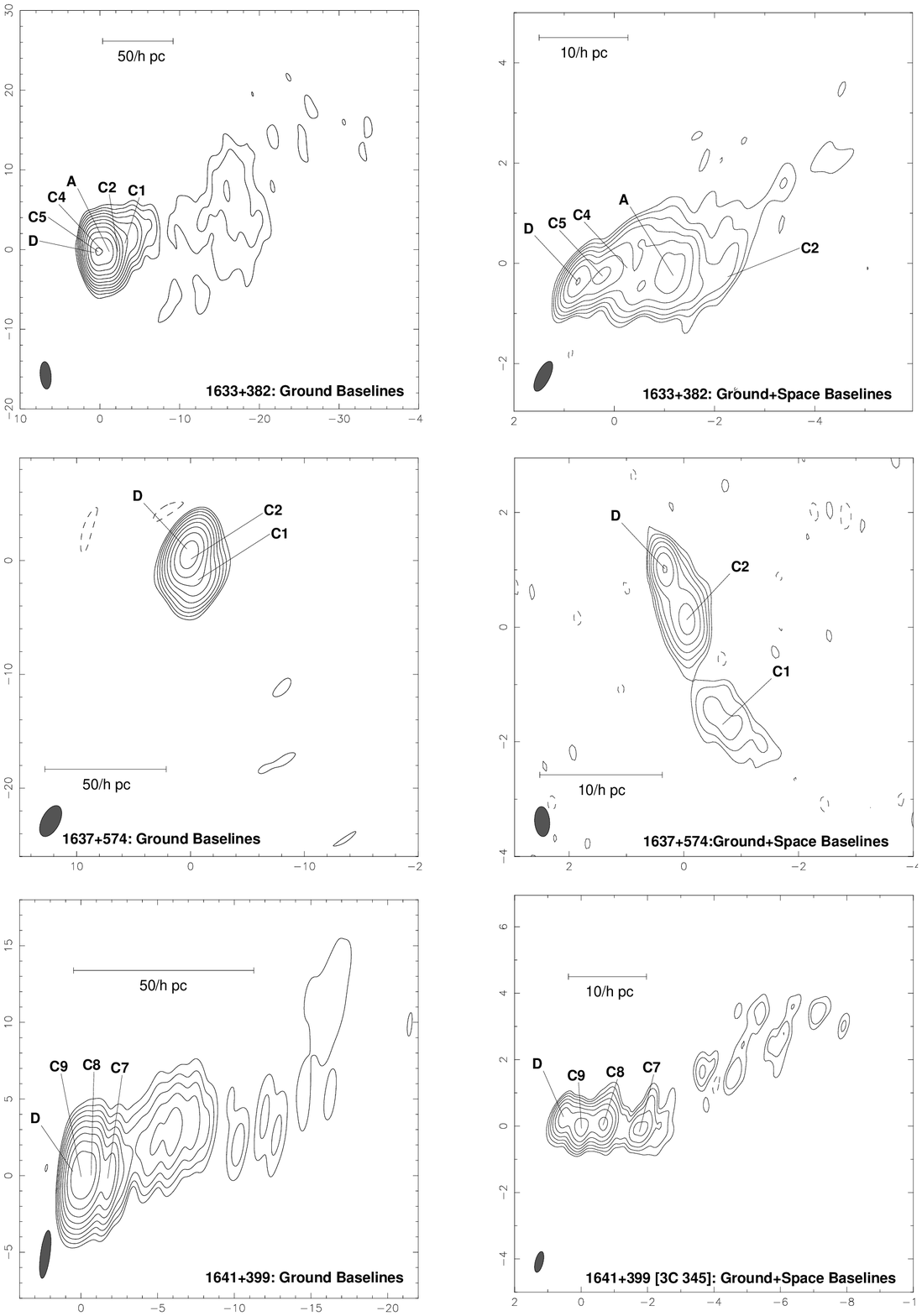}
\caption{\small Left column: naturally-weighted images using
ground-to-ground baselines only. Right column: uniformly-weighted images using
all baselines. Angular scales are in milliarcseconds. }
\end{figure}

\begin{figure}
\plotone{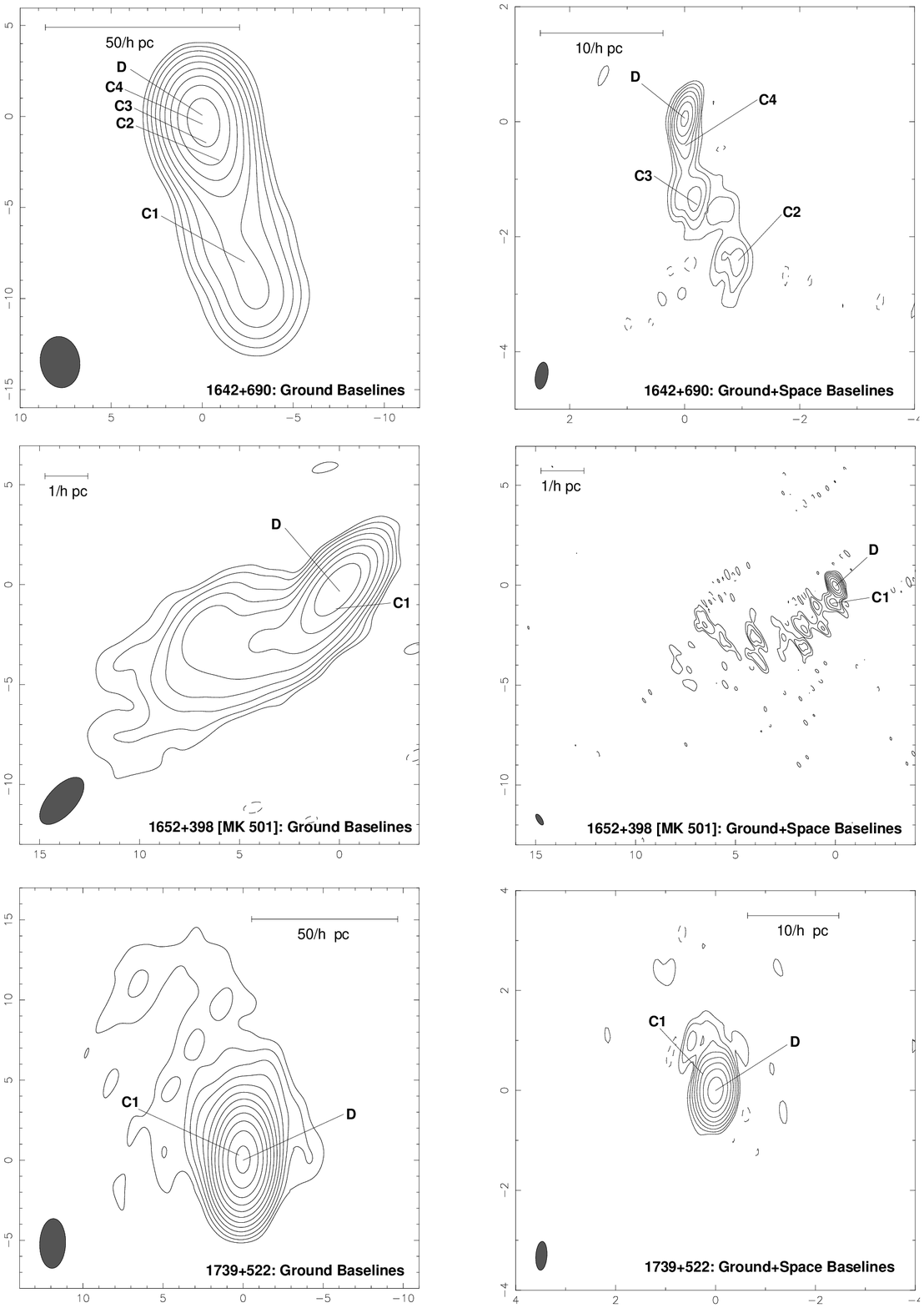}
\caption{\small Left column: naturally-weighted images using
ground-to-ground baselines only. Right column: uniformly-weighted images using
all baselines. Angular scales are in milliarcseconds. }
\end{figure}

\begin{figure}
\plotone{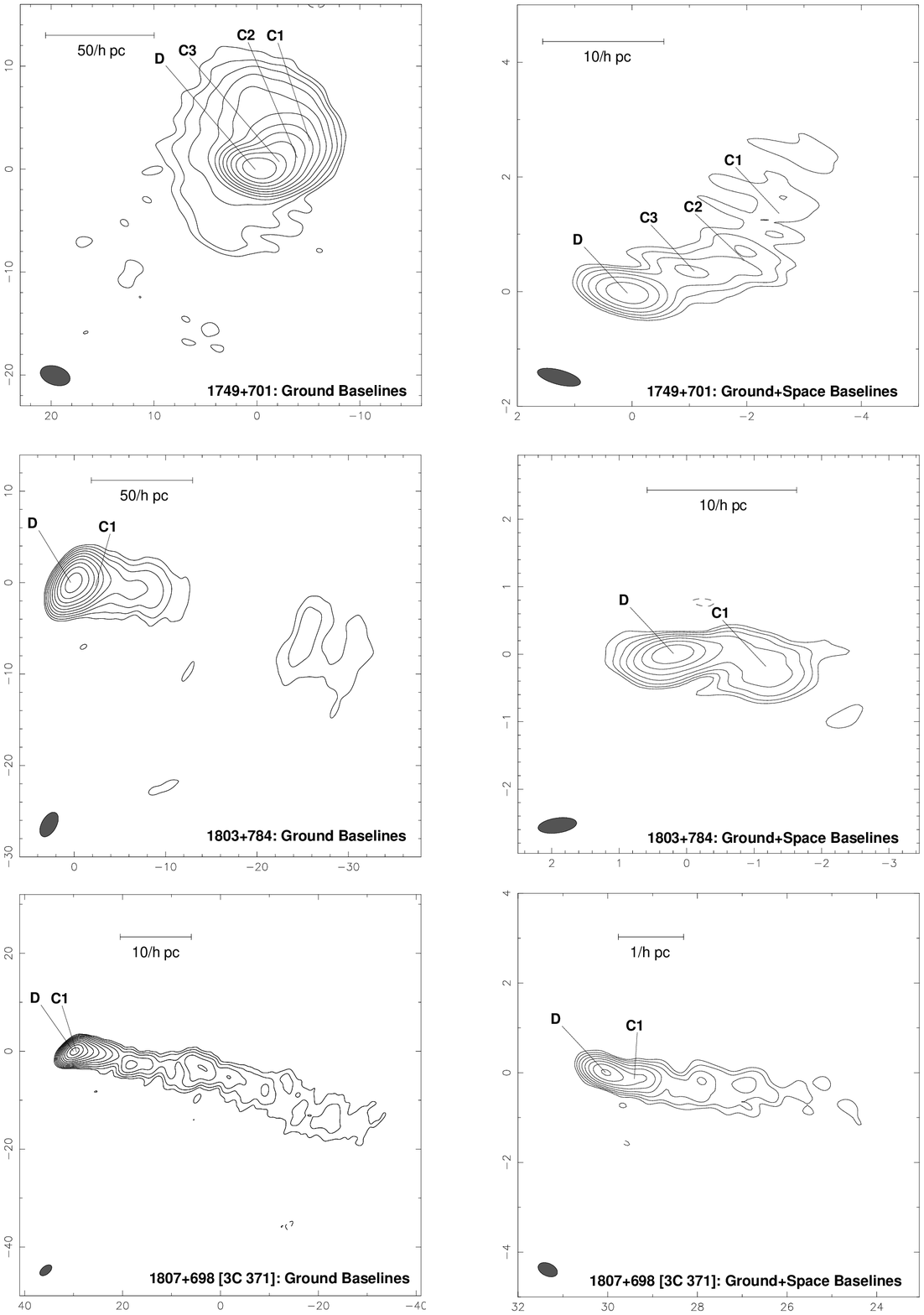}
\caption{\small Left column: naturally-weighted images using
ground-to-ground baselines only. Right column: uniformly-weighted images using
all baselines. Angular scales are in milliarcseconds. }
\end{figure}

\begin{figure}
\plotone{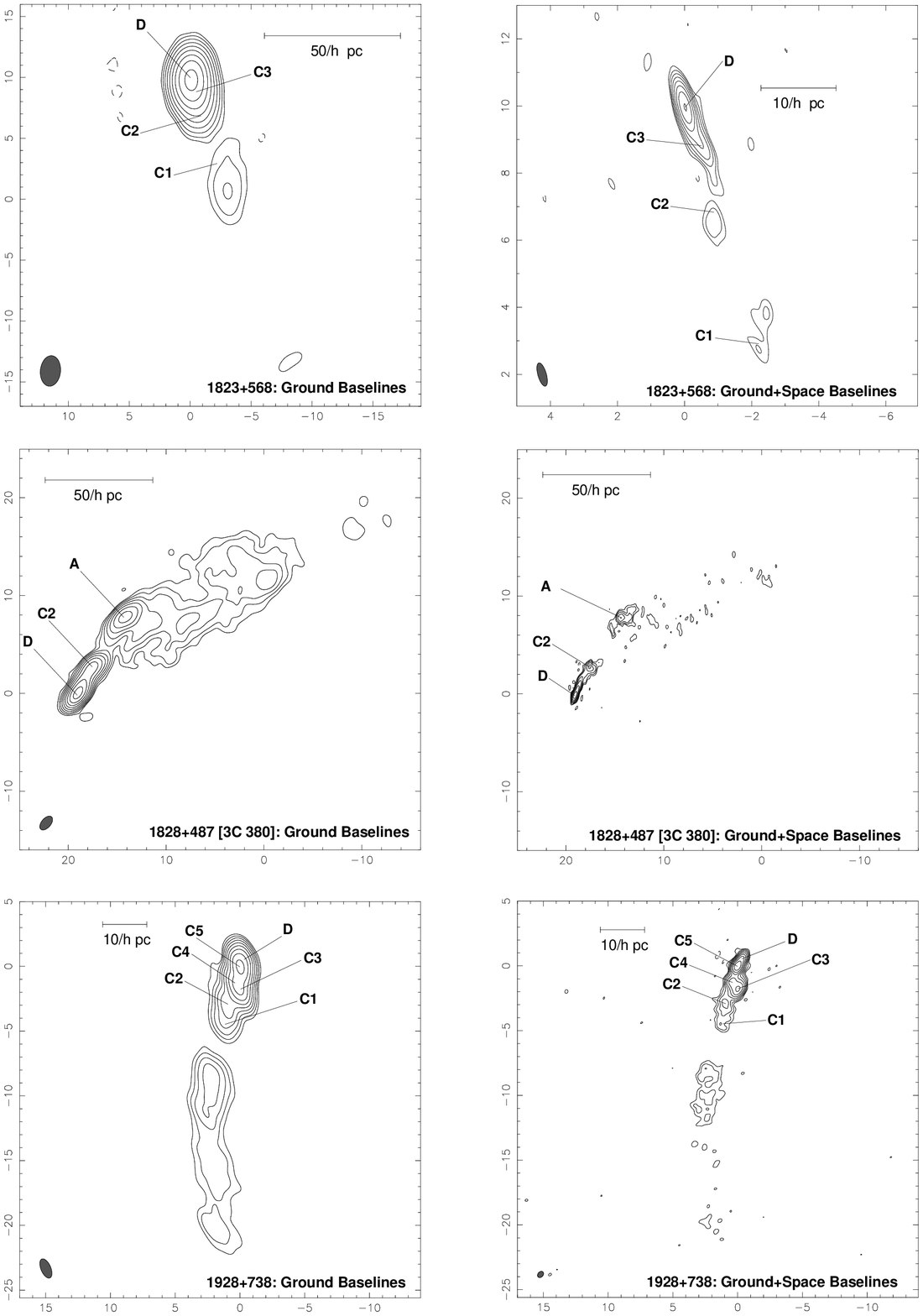}
\caption{\small Left column: naturally-weighted images using
ground-to-ground baselines only. Right column: uniformly-weighted images using
all baselines. Angular scales are in milliarcseconds. }
\end{figure}

\begin{figure}
\plotone{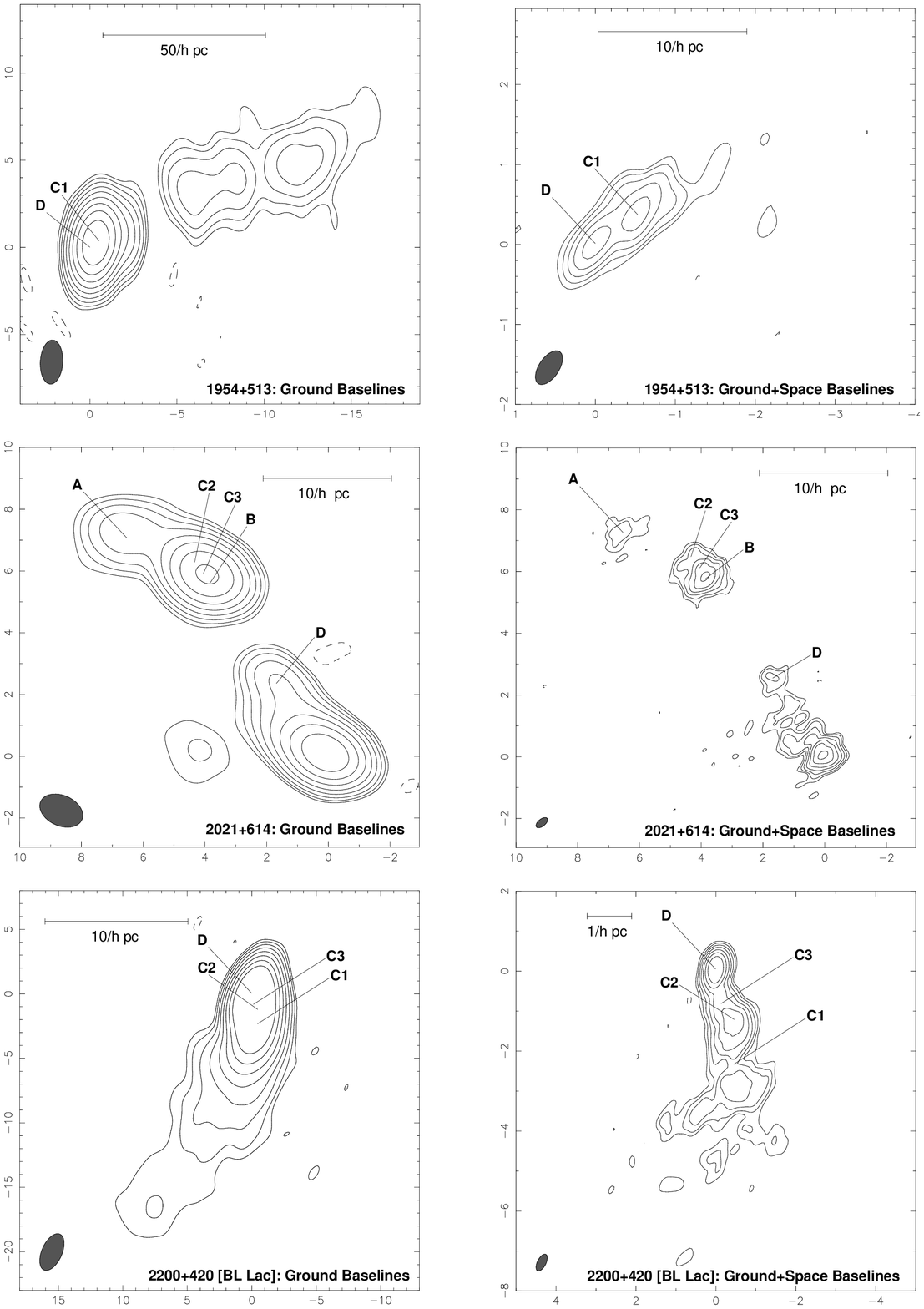}
\caption{\small Left column: naturally-weighted images using
ground-to-ground baselines only. Right column: uniformly-weighted images using
all baselines. Angular scales are in milliarcseconds. }
\end{figure}

\begin{figure*}
\epsscale{1}
\plottwo{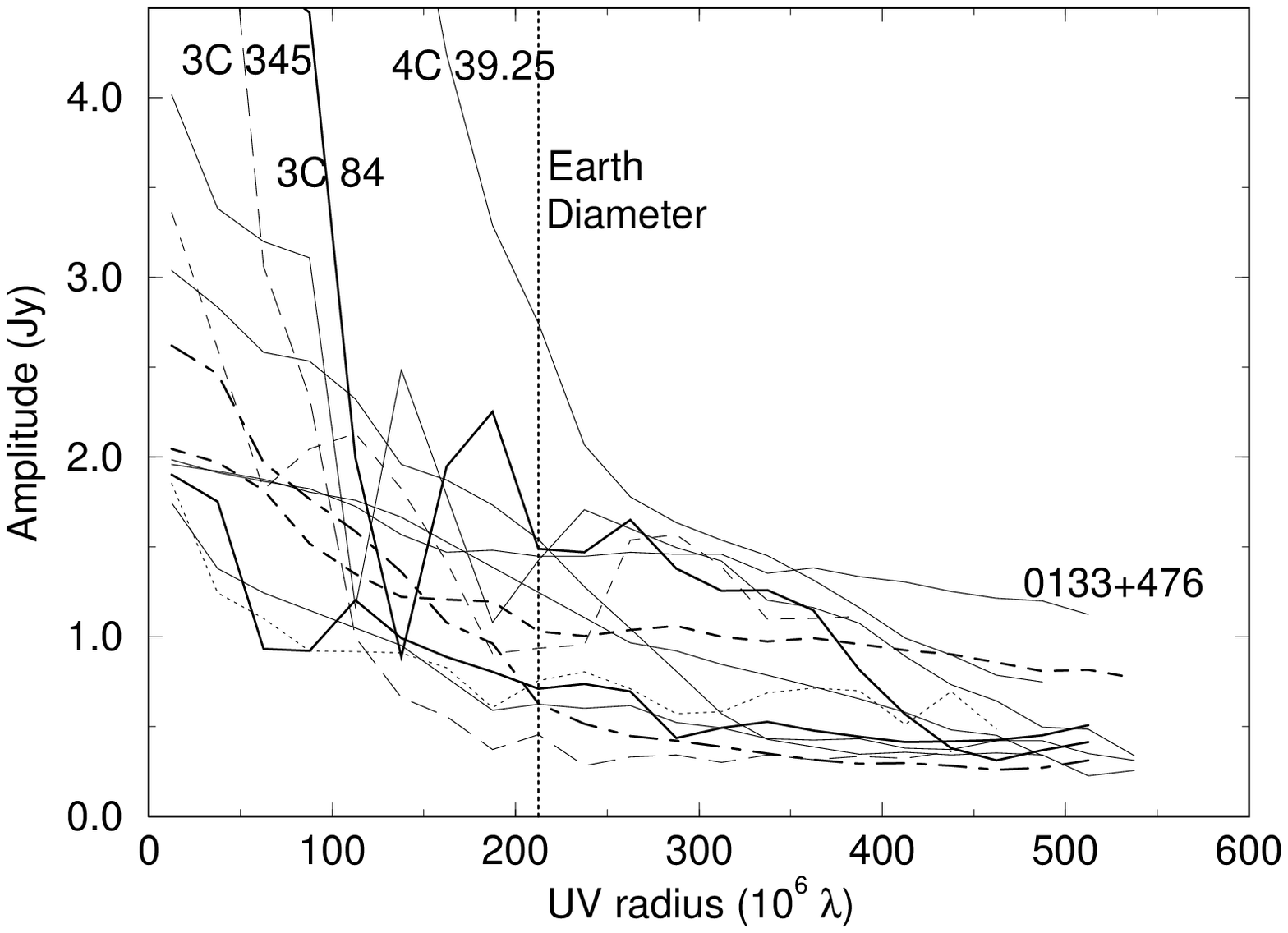}{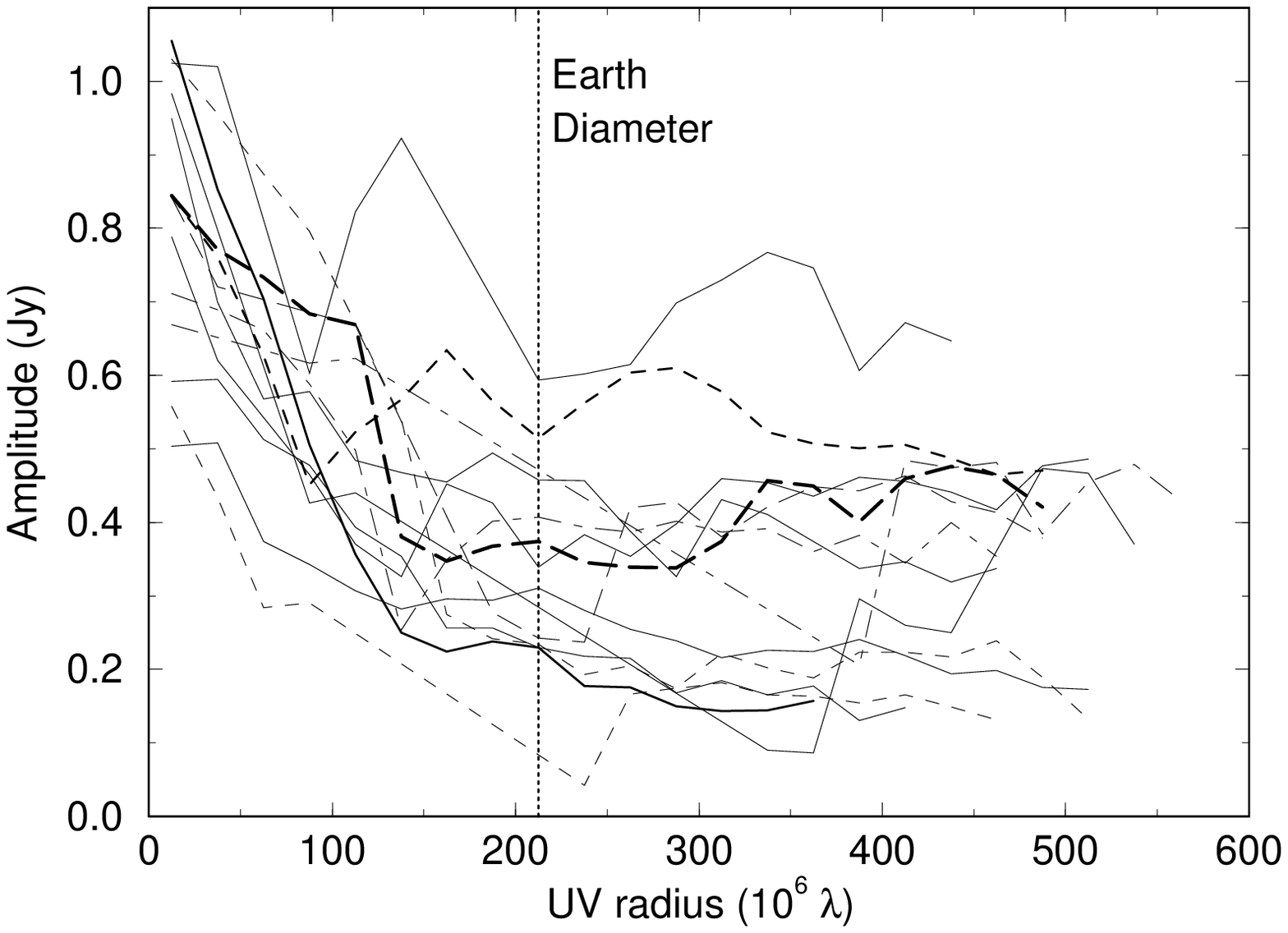}
\caption{\label{envelope} Upper envelope of visibility amplitude distribution plotted
against projected baseline separation for strong (left panel) and weak
(right panel) sources in Pearson-Readhead sample. The flattening on
baselines longer than an Earth diameter is indicative of strong
unresolved core components. }
\end{figure*}

\begin{figure*}
\plotone{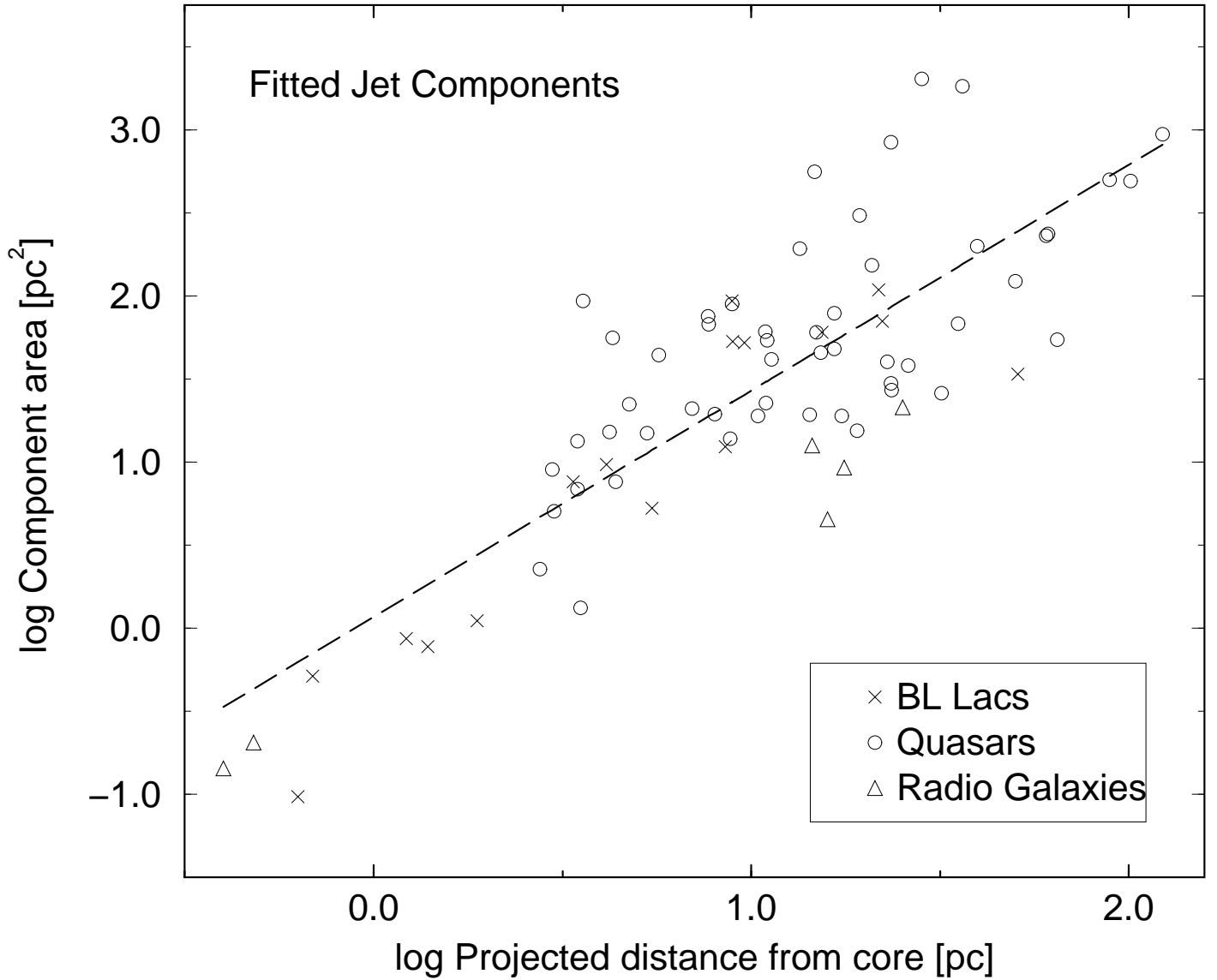}
\caption{\label{r__areacpt} Fitted Gaussian component size plotted
against projected separation from the core component, for jets in
the Pearson-Readhead survey.}
\end{figure*}

\begin{figure*}
\plottwo{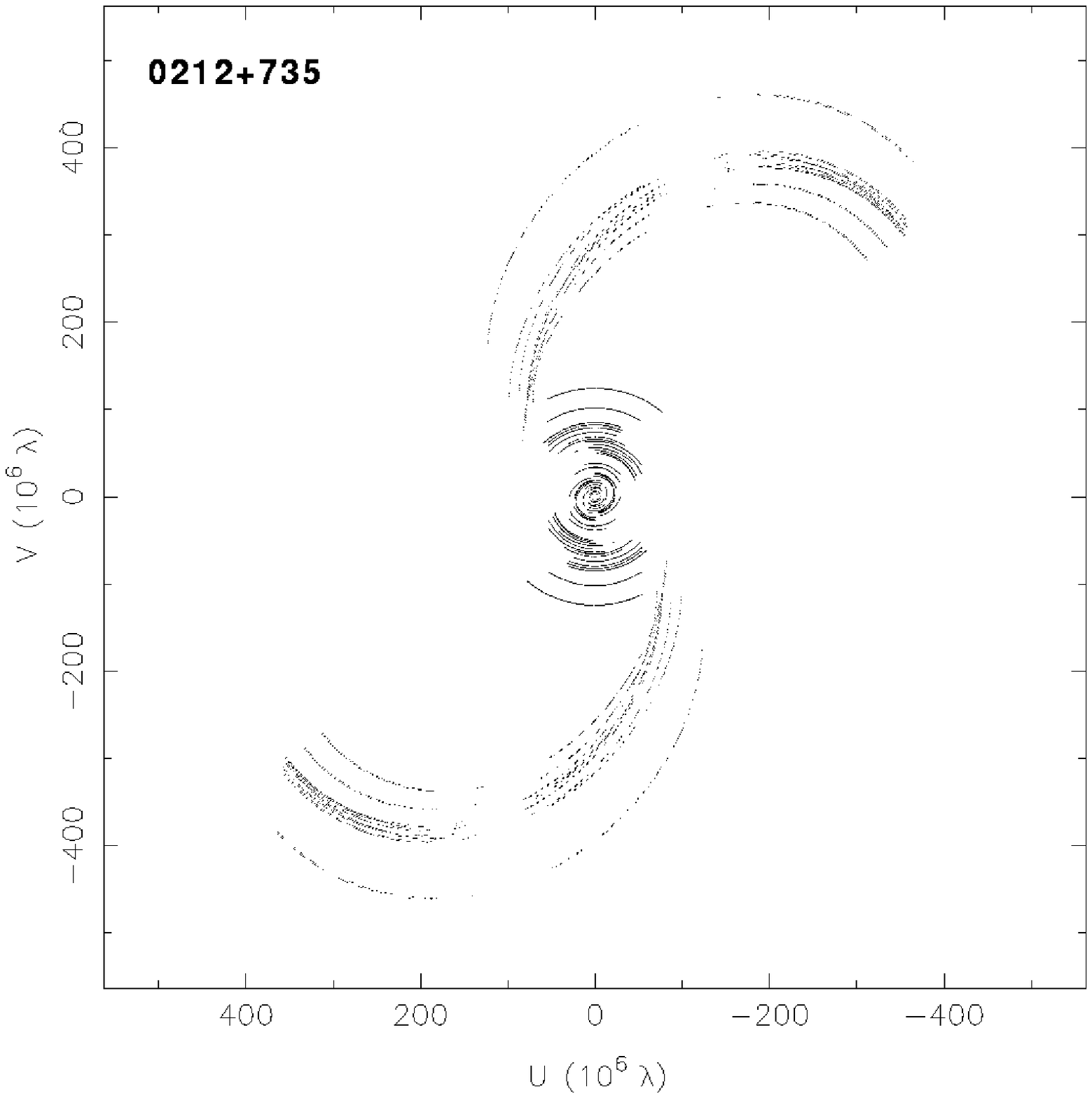}{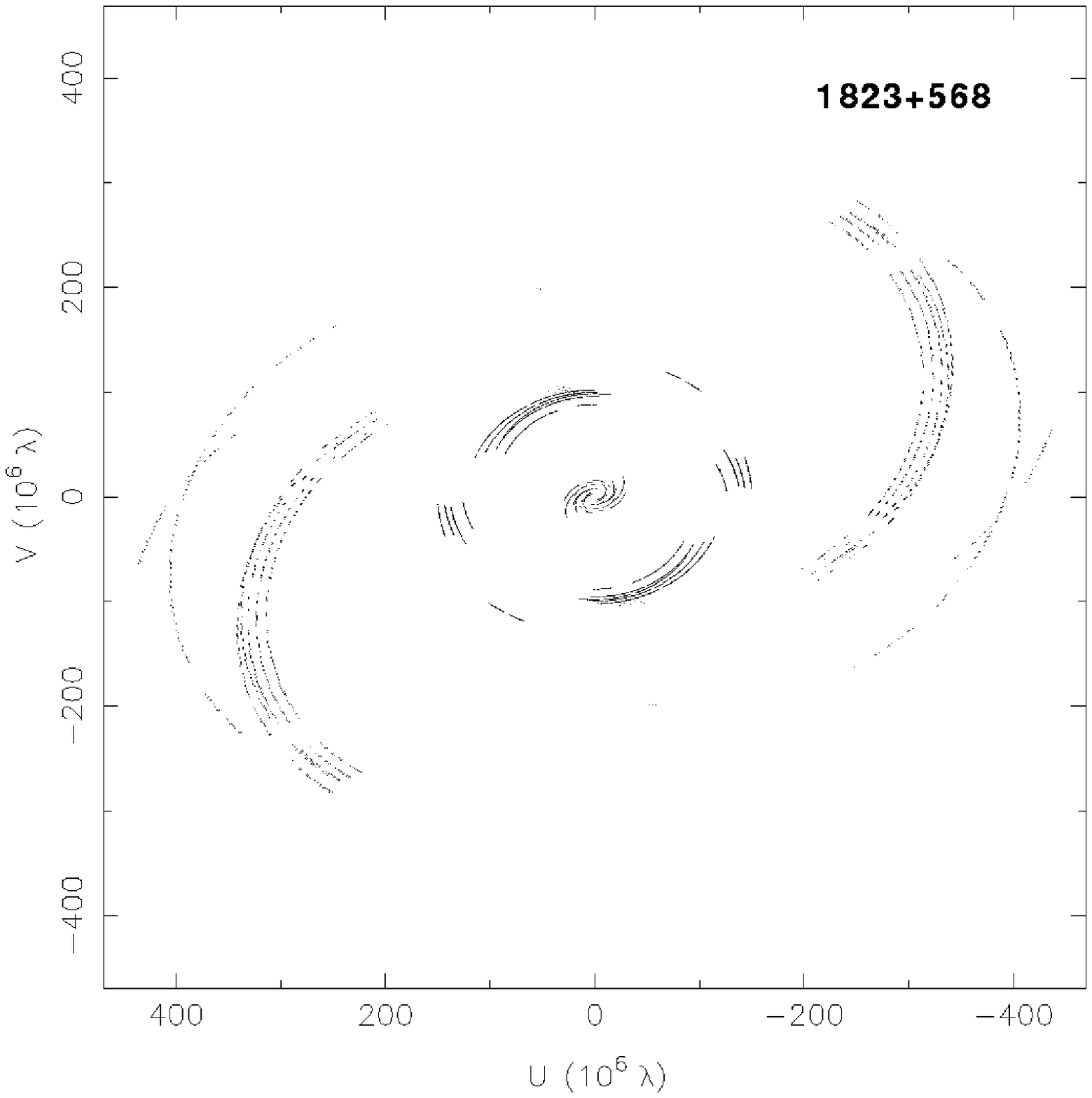}
\caption{\label{uvplot} Plots showing typical $(u,v)$ plane coverages for
VSOP observations using HALCA and the VLBA (left panel) and the EVN
(right panel). }
\end{figure*}

\begin{figure}[htbp]
\includegraphics[scale=0.75]{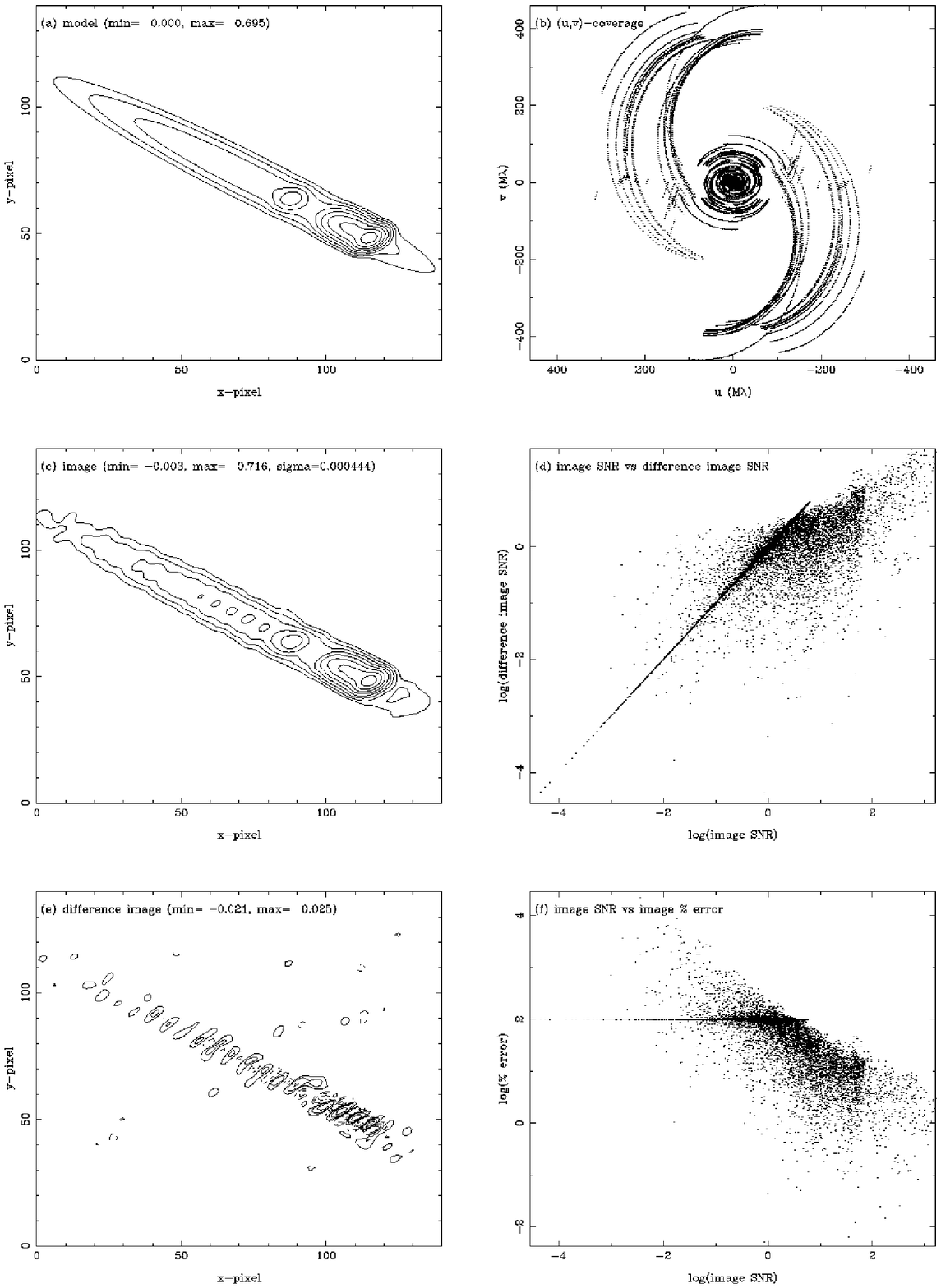}
\caption{\scriptsize Simulations using complex core-jet source model
({\it cj-complex})
and jet PA = 60$^{\circ}$.  The images have a pixel size of 0.08
milliarcseconds. Panels are as follows: (a) original source model; (b)
$(u,v)$ coverage; (c) simulated HALCA image; (d) SNR ratio of each
pixel in the difference image versus SNR in simulated image; (e)
difference image; (f) Percentage error of each pixel in simulated
image plotted against SNR ratio. The contours in panels (a) and (c)
are 0.5,1,2,4,8,16,32, and $64\%$ of the peak for both positive and
negative contours. The contours in panel (e) are 8,16,32, and $64\%$
of the peak for both the positive and negative contours. The pixels
clustered along the straight lines in panels (d) and (f) are located
mainly off-source and have 100$\%$ error, since the model is
essentially zero at these points. \label{fig_diffex1}}
\end{figure}

\begin{figure}[htbp]
\includegraphics[scale=0.75]{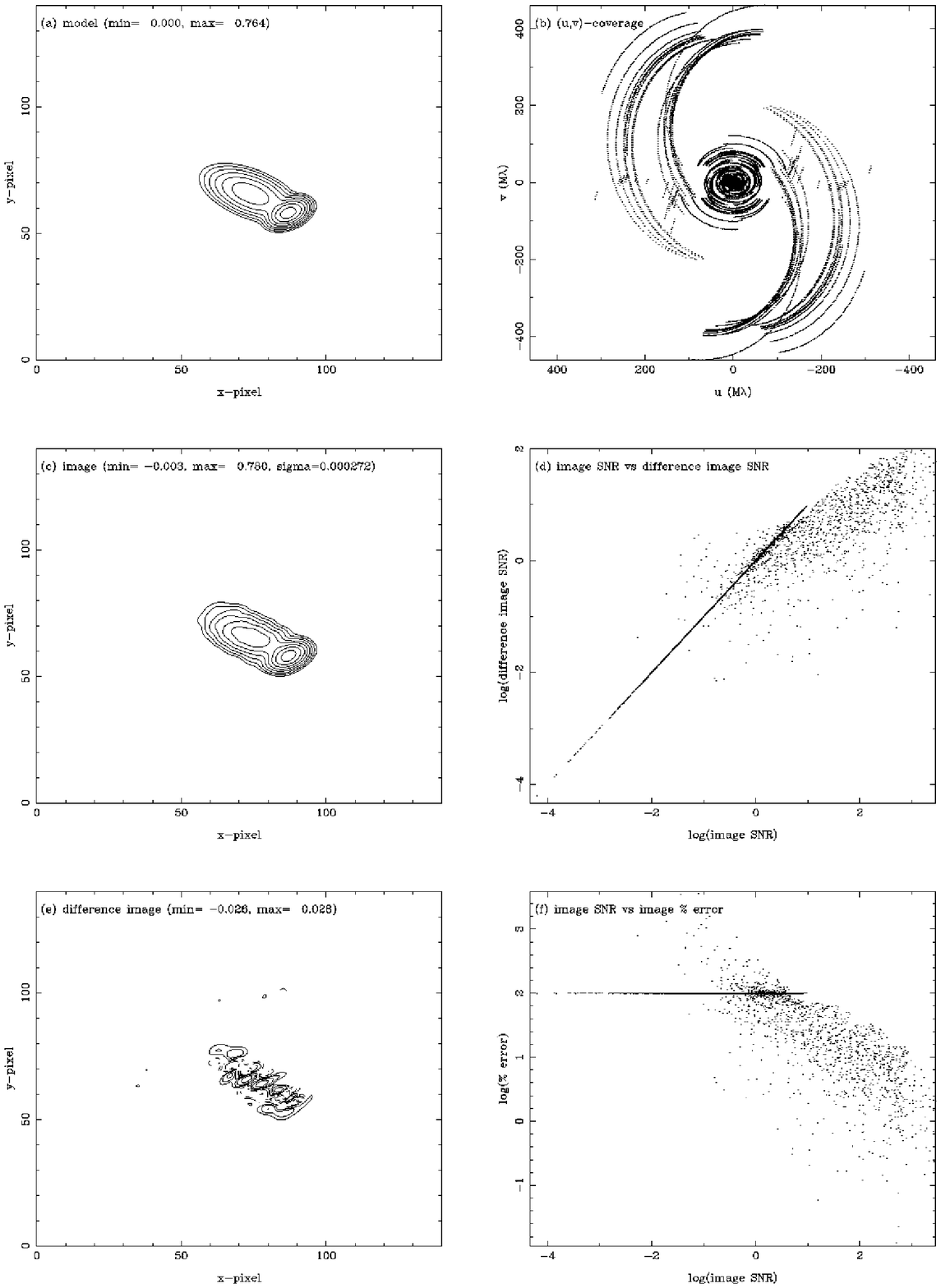}
\caption{\scriptsize Simulations using simple core-jet source model
({\it cj-simple})
and jet PA = 60$^{\circ}$.  The images have a pixel size of 0.08
milliarcseconds. Panels are as follows: (a) original source model; (b)
$(u,v)$ coverage; (c) simulated HALCA image; (d) SNR ratio of each
pixel in the difference image versus SNR in simulated image; (e)
difference image; (f) Percentage error of each pixel in simulated
image plotted against SNR ratio. The contours in panels (a) and (c)
are 0.5,1,2,4,8,16,32, and $64\%$ of the peak for both positive and
negative contours. The contours in panel (e) are 8,16,32, and $64\%$
of the peak for both the positive and negative contours. The pixels
clustered along the straight lines in panels (d) and (f) are located
mainly off-source and have 100$\%$ error, since the model is
essentially zero at these points. \label{fig_diffex2}}
\end{figure}

\begin{figure}[htbp]
\includegraphics[angle=270, scale=0.75]{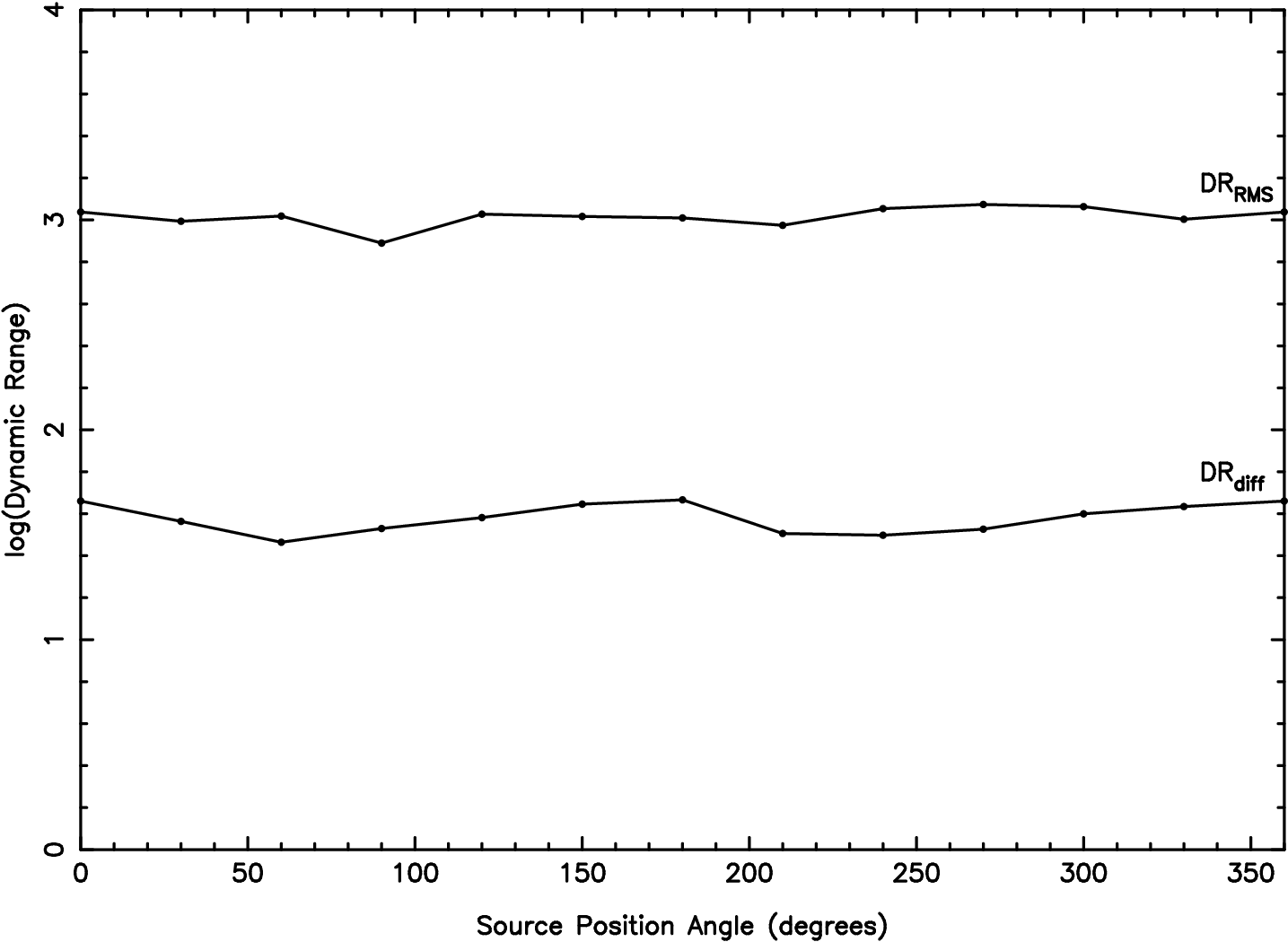}
\caption{ Dynamic range as a function of source position angle (PA) for
the complex 
core-jet source model ({\it cj-complex}). The upper curve (labeled $DR_{\rm
RMS}$) represents the ratio of maximum flux density in the image to
the off-source RMS noise level. The lower curve (labeled $DR_{\rm
diff}$) represents the maximum in the image divided by the
maximum value (whether positive or negative) in the difference
image. The major axis of the $(u,v)$ coverage is held constant at PA=
0$^\circ$ for all models. 
\label{fig_pa1}}
\end{figure}

\begin{figure}[htbp]
\includegraphics[angle=270, scale=0.75]{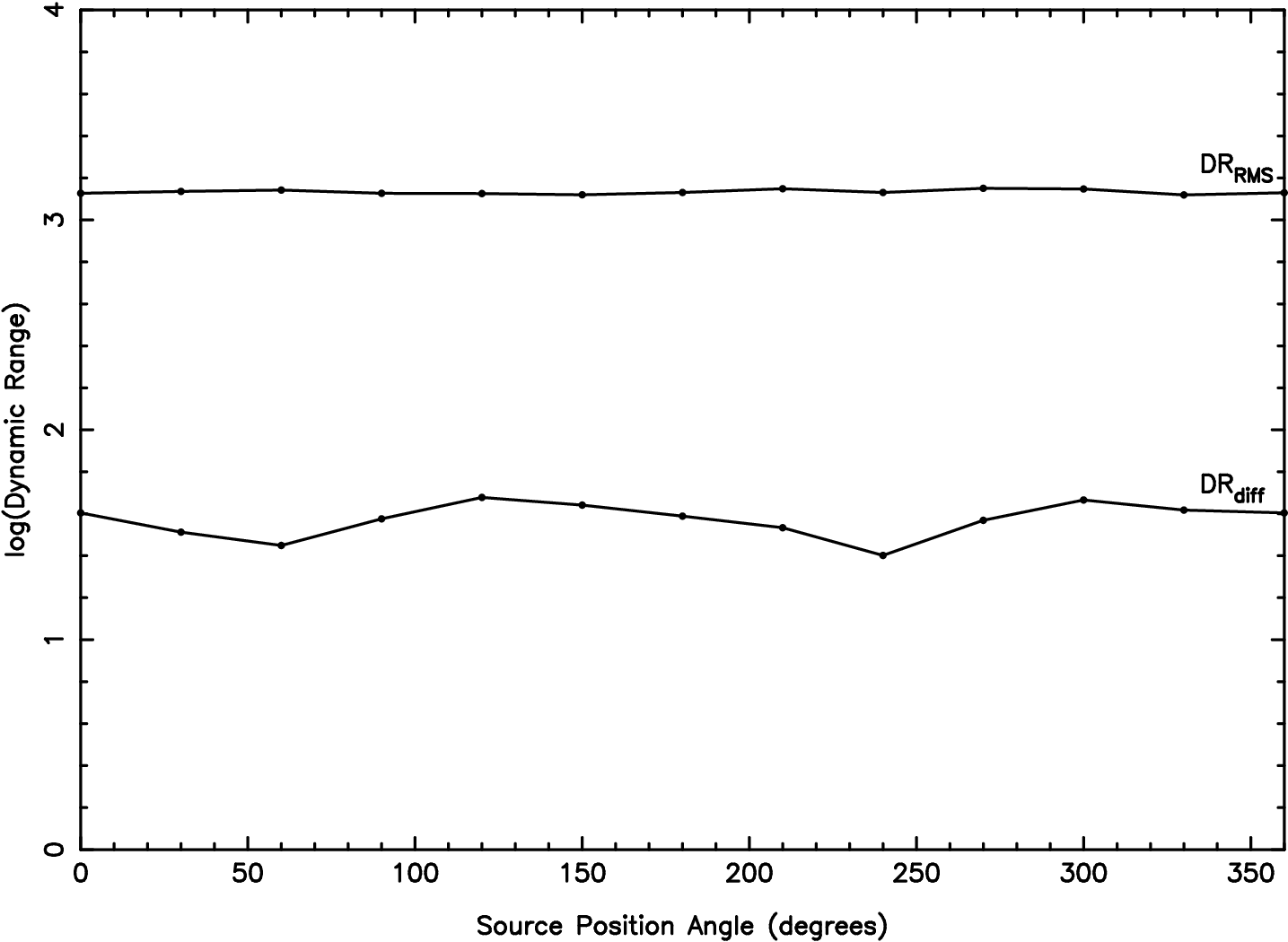}
\caption{Dynamic range as a function of source position angle (PA) for the
simple core-jet source model ({\it cj-simple}). The upper curve (labeled $DR_{\rm
RMS}$) represents the ratio of maximum flux density in the image to
the off-source RMS noise level. The lower curve (labeled $DR_{\rm
diff}$) represents the maximum in the image divided by the
maximum value (whether positive or negative) in the difference
image. The major axis of the $(u,v)$ coverage is held constant at PA=
0$^\circ$ for all models.
\label{fig_pa2}}
\end{figure}

\end{document}